\documentclass[10pt,a4paper]{article}

\usepackage[utf8]{inputenc}
\usepackage{amsmath,amsfonts,amssymb,amsthm,hyperref}
\usepackage{nicefrac}
\usepackage{graphicx}
\graphicspath{{figures/}}
\DeclareGraphicsExtensions{.pdf}

\newcommand{\norm}[1]{\left\lVert#1\right\rVert}
\newcommand\abs[1]{\left|#1\right|}

\newtheorem{remark}{Remark}

\title{A two-dimensional data-driven model for traffic flow on highways}
\date{\today}
\author{Michael Herty \\
		{\small\it Institut f\"{u}r Geometrie und Praktische Mathematik (IGPM)} \\
		{\small\it RWTH Aachen University} \\
		{\small\it Templergraben 55, 52062 Aachen, Germany\vspace{5mm}} \\
	Adrian Fazekas\\
		{\small\it Institut f\"{u}r Stra{\ss}enwesen (ISAC)} \\
		{\small\it RWTH Aachen University} \\
		{\small\it Mies-van-der-Rohe-Str. 1, 52074 Aachen, Germany\vspace{5mm}} \\
	Giuseppe Visconti \\
		{\small\it Institut f\"{u}r Geometrie und Praktische Mathematik (IGPM)} \\
		{\small\it RWTH Aachen University} \\
		{\small\it Templergraben 55, 52062 Aachen, Germany}
	   }

\begin{document}
\maketitle    
\begin{abstract}
	Based on experimental traffic data obtained from German and US highways, we propose
	a novel two-dimensional first-order macroscopic traffic flow model. The goal is to reproduce a detailed description of traffic dynamics for the real road geometry. In our approach both the dynamics along the road and across the lanes is continuous. The closure relations, being necessary to complete the hydrodynamics equation, are obtained by regression on fundamental diagram data. Comparison with prediction of one-dimensional models shows the improvement in performance of the novel model.
\end{abstract}



\section{Introduction} \label{sec:Introduction}

The mathematical modeling of vehicular traffic flow uses different descriptions and we refer to 
\cite{BellomoDogbe2011,Helbing2001,piccoli2009ENCYCLOPEDIA} for some review papers. Besides microscopic and cellular 
models there has been intense research in continuum models where the temporal and spatial evolution 
of car densities is prescribed. Based on the level of detail there are gas-kinetic or mesoscopic models (e.g., \cite{HertyIllner2010aa,HertyPareschi2010ab,IllnerKlarMaterne2003,KlarWegener2000ab,Phillips1979,HermanPrigogine1971aa}) and macroscopic models being fluid-dynamics models (e.g., \cite{aw2000SIAP,BayenClaudel2011aa,BerthelinDegondDelitalaRascle2008ab,Daganzo1995aa, Daganzo2006,GaravelloPiccoli2006ab,Goatin2006aa, KernerKonhauser1993,KernerKonhauser1994,Lebacque1993aa,lighthill1955PRSL,Payne1971aa,Payne1979aa,richards1956OR,Underwood1961,ZhangMacro}). Among the (inviscid) macroscopic models one typically distinguishes between first-order models based on scalar hyperbolic equations and second-order models comprised of systems of hyperbolic equations. The pioneering work of the first case is the Lighthill and Whitham \cite{lighthill1955PRSL} and Richards \cite{richards1956OR} model (LWR). While a specific example of the second case is the Aw and Rascle \cite{aw2000SIAP} and Zhang \cite{ZhangMacro} model (ARZ).
Depending on the detailed level of description of the underlying process different models have been employed and tested against data. In recent publications it has been argued that the macroscopic models provide a suitable framework for the incorporation of on-line traffic data and in particular of fundamental diagram data \cite{MobileMillennium,BayenClaudel2010aa,FanHertySeibold}. While microscopic models are nowadays widely used in traffic engineering, continuum models have been studied mathematically, but very little work has been conducted on their validation with traffic data~\cite{MobileMillennium,Aminetal2008,BlandinBrettiCutoloPiccoli2009aa}.

So far, most of the proposed continuum models are for single lane vehicular traffic dynamics. However, the data for fundamental diagrams is taken from interstate roads with multiple lanes \cite{NGSIM,RTMC} and can be used for deriving or testing models for real road geometry. Multi-lane models belong to this class. Typical modeling of multi-lane traffic uses a spatially one-dimensional model (1D) of either LWR or ARZ type for each lane. The lane-changing of cars is then modeled by interaction terms (the sources on the right-hand sides of the
equations) using empirical interaction rates, see e.g. \cite{klar1999SIAP-1,klar1999SIAP-2,KlarWegener2000ab}. The interaction modeling is typically assumed to be proportional to local density on current and desired lane. A fluid dynamics model describing the cumulative density on all lanes is proposed in \cite{ChetverushkinChurbanovaFurmanovTrapeznikova2010,ChetverushkinChurbanovaSukhinovaTrapeznikova2008,SukhinovaTrapeznikovaChetverushkinChurbanova2009}, where a two-dimensional (2D) system of balance laws is obtained by analogy with the quasi-gas-dynamics (QGD) theory. Here, the authors model 2D dynamics assuming that vehicles move to lanes with a faster speed or a lower density and the evolution equation for the lateral velocity is described by the sum of the three terms proportional local density and mean speed along the road.

A major problem of the approaches described above is to estimate from data the interaction rate or the great number of coefficients and parameters. Therefore, here, we propose a different approach: we treat also lanes as continuum and postulate a dynamics orthogonal to the driving direction. The precise form of the dynamics is established through comparisons with fundamental diagrams obtained from trajectory data recorded on a road section of the A3 German highway near Aschaffenburg. Thus, the experimental measurements allow us to derive a model being able to take into account the realistic dynamics on the real road geometry without prescribing heuristically the behavior of the flow of vehicles.

The contribution and the organization of this paper is summarized below.
\begin{itemize}
	\item[(i)] Derivation and the presentation of historic fundamental diagrams data  for the dynamics of traffic across the lanes (see Section \ref{sec:DataSet}). In fact, the German data-set provides the two-dimensional time-dependent positions of vehicles while crossing the road section. Therefore, in addition to the classical fundamental diagrams widely studied in the literature \cite{GreenshieldsSymposium,kerner2004BOOK,Leutzbach} and used for deriving one-dimensional data-fitted macroscopic models \cite{FanHertySeibold,FanSeibold2012}, we can also generate diagrams for the dynamics across the lanes. Although two-dimensional experimental traffic measurements are already available in the literature, this is, to our knowledge, the first time that they are used to study the dynamics orthogonal to the movement of vehicles;
	\item[(ii)] Design of a new data-fitted two-dimensional first-order model and the analysis of its mathematical properties (see Section \ref{sec:MacroModel}). The historic data are therefore used to develop the novel macroscopic model defining the flux functions by means of a data-fitting approach. The closures are necessary to complete the macroscopic equation and taking them using the experimental data allows to describe the real dynamics of the flow;
	\item[(iii)] validation of the novel 2D macroscopic model via time-dependent trajectory data and the definition of a systematic methodology to study and to compare the predictive accuracy with respect to the 1D LWR model (see Section \ref{sec:Simulations}).
\end{itemize}

Finally, we end the paper with a concluding part (see Section \ref{sec:Conclusions}) dealing with final comments and perspectives. In particular, we briefly discuss on the difference between the German data-set and US data, e.g. \cite{NGSIM,RTMC}, since the latter provide a naive behavior of the flow across the lanes.

\section{Data-set description and fundamental diagrams} \label{sec:DataSet}
%mh

We use a set of experimental data recorded on a German highway. Precisely, we have two-dimensional trajectory data collected on a $80$~meter stretch of the westbound direction of A3~highway near Aschaffenburg. Laser scanners detect the two-dimensional positions $\left(x_i(t),y_i(t)\right)$ of each vehicle $i$ at time $t$ on the road segment with a temporal resolution of $0.2$~seconds for a total time of approximately $20$~minutes. Here the position $x$ is in driving direction, the position $y$ is across lanes. During the time observation, the laser scanners record the trajectories of $1290$ vehicles.

The road section consists of three lanes and an outgoing ramp. However, we only consider the stretch as if there is no ramp. In fact, the data show that the flow on the ramp does not influence the traffic conditions, namely the amount of traffic on the ramp is not significant. Taking into account only the three main lanes, the road width is $12$ meter. The stretch we are considering is not straight but it has a turn with a small radius of curvature. We have taken into account this feature when cleaning the experimental data since the curvature could mainly falsify the data in $y$-direction. Precisely, the cleaning procedure is based on the knowledge of the curvature of the road-section. We clean the time-dependent $y$-positions of the vehicles on the road by adjusting them by a factor which allows to not consider movement in $y$ if it is only due to the curvature of the road section.

As pointed out in the Introduction, in this paper we are interested in the study of macroscopic traffic models. In other words, instead of looking at the motion of each single vehicle, we wish to ``zoom out'' to a more aggregate level by treating traffic as a fluid. Therefore, with the aim of proposing a novel data-fitted 2D macroscopic model, from the microscopic experimental data we need to recover the macroscopic quantities, namely the \emph{density} (measured as number of vehicles per kilometer), the \emph{flux} (measured as number of vehicles per hour) and the \emph{mean speed} (measured as kilometer per hour) of the flow.

To this end, firstly, we observe that the microscopic positions $\left(x_i(t),y_i(t)\right)$ of vehicles at each time are sufficient to recover the microscopic velocities of vehicles. In particular, since the road section is relatively short, we compute the velocity, both in $x$- and $y$-direction, of each vehicle by using a linear approximation in the least squares sense of its positions, $x_i(t)$ and $y_i(t)$ respectively, on the road during the time interval. In other words we assume that the vehicle velocity is constant during the crossing of the road section and is exactly the slope of the linear fit. Thus, we associate at each vehicle $i$ the vector of the microscopic velocities $\left(v^x_i,v^y_i\right)$. The maximum detected speed in $x$-direction is about $120$ kilometer per hour which means about $2.7$ seconds to travel the $80$ meters of the road section.

The time-dependent microscopic positions $\left(x_i(t),y_i(t)\right)$ and the microscopic velocities $\left(v^x_i,v^y_i\right)$ of vehicles are used to compute the macroscopic data as we describe in the following. Clearly, since we are aimed to develop a two-dimensional data-fitted first-order macroscopic model, the derivation of macroscopic quantities, such as the flux and the mean speed, should be done for each direction, along the road ($x$-direction) and across the lanes ($y$-direction), separately.

The macroscopic density gives information on the congestion level of the road section. It is usually expressed in number of vehicles per unit length (here kilometers) and therefore it ignores the concept of traffic composition. This is not restrictive for our purpose of deriving a two-dimensional first-order macroscopic model for traffic. The modeling of the heterogeneous composition of vehicles is studied in multi-population models, e.g., in~\cite{Benzoni-GavageColombo2003ab} at the macroscopic level and in~\cite{PgSmTaVg3,PgSmTaVg} at the kinetic level, where the concept of density is replaced by the \emph{rate of occupancy}. In order to compute the macroscopic density, we first fix a sequence of $M+1$ equally spaced discrete times $\{t_k\}_{k=0}^M$ such that $t_{k+1}-t_k=dt$, $t_0=0$ and $t_M=t_{\max}$, where $t_{\max}$ is the final observation time in the data-set (here 20 minutes). Then we count the number of vehicles $N(t_k)$ on the road at each discrete time $t_k$ defining
\[
\tilde{\rho}(t_k) = \frac{N(t_k)}{L}, \quad k=0,\dots,M
\]
where $L$ is the length of the road section expressed in the unit length. Finally, we consider a moving mean by aggregating with respect a certain time period $T$, with $T \ll t_{\max}$ and including $m$ consecutive observations. This temporal average leads to $\left\lceil \frac{M+1}{m} \right\rceil+1$ values of the density
\[
\rho_{k_0} = \frac{1}{T} \sum_{k=k_0}^{k_0+m-1} \tilde{\rho}(t_k), \quad k_0 = 0,\dots,\left\lceil \frac{M+1}{m} \right\rceil.
\]
In particular, in this paper we take $dt=1$ second and then we aggregate the data over the time period $T=60$ seconds.

Observe that, clearly, the density does not depend on the direction we are looking at. The computation of the flux and of the mean speed is, instead, a little bit more complex.

In our approach, we first compute the mean speeds of the flow. Consider the same sequence of $M+1$ equally spaced discrete times $\{t_k\}_{k=0}^M$ introduced above. Then we average the microscopic velocities $v_i^x$ and $v_i^y$ of all vehicles being on the road at a fixed time $t_k$ with respect to the number of vehicles $N(t_k)$, so that we define
\[
\tilde{u}^x(t_k) = \frac{1}{N(t_k)} \sum_{i=1}^{N(t_k)} v^x_i, \quad \tilde{u}^y(t_k) = \frac{1}{N(t_k)} \sum_{i=1}^{N(t_k)} v^y_i, \quad k=0,\dots,M.
\]
Using $\tilde{u}^x(t_k)$ and $\tilde{u}^y(t_k)$ we then compute the fluxes at each discrete time $t_k$ by means of the hydrodynamics relation
\[
\tilde{q}^x(t_k) = \tilde{\rho}(t_k) \tilde{u}^x(t_k), \quad \tilde{q}^y(t_k) = \tilde{\rho}(t_k) \tilde{u}^y(t_k), \quad k=0,\dots,M
\]
and, as done for the density, we consider a temporal average by aggregating with respect the same time period $T$, leading to the following expressions of the macroscopic fluxes
\begin{align*}
q^x_{k_0} & = \frac{1}{T} \sum_{k=k_0}^{k_0+m-1} \tilde{q}^x(t_k) = \frac{1}{TL} \sum_{k=k_0}^{k_0+m-1} \sum_{i=1}^{N(t_k)} v^x_i, \\ q^y_{k_0} &= \frac{1}{T} \sum_{k=k_0}^{k_0+m-1} \tilde{q}^y(t_k) = \frac{1}{TL} \sum_{k=k_0}^{k_0+m-1} \sum_{i=1}^{N(t_k)} v^y_i,
\end{align*}
for $k_0=0,\dots,\left\lceil \frac{M+1}{m} \right\rceil$. Finally, using again the hydrodynamics relation, we get the mean speeds of the flow as
\[
u^x_{k_0} = \frac{q^x_{k_0}}{\rho_{k_0}}, \quad u^y_{k_0} = \frac{q^y_{k_0}}{\rho_{k_0}}, \quad k_0=1,\dots,\left\lceil \frac{M+1}{m} \right\rceil.
\] 

For a more detailed discussion on the computation of macroscopic quantities from microscopic data, we refer to~\cite{Hoogendoorn2007,MaerivoetDeMoor}.

The diagrams showing the relations between the vehicle density $\rho$ and the fluxes $q^x$, $q^y$ or the mean speeds $u^x$, $u^y$ are called \emph{fundamental diagrams} and \emph{speed-density diagrams}, respectively. They represent the basic tools for the analysis of traffic problems operating in a homogeneous \emph{steady state} or \emph{equilibrium} conditions.

\begin{figure}[t!]
	\centering
	\includegraphics[width=0.49\textwidth]{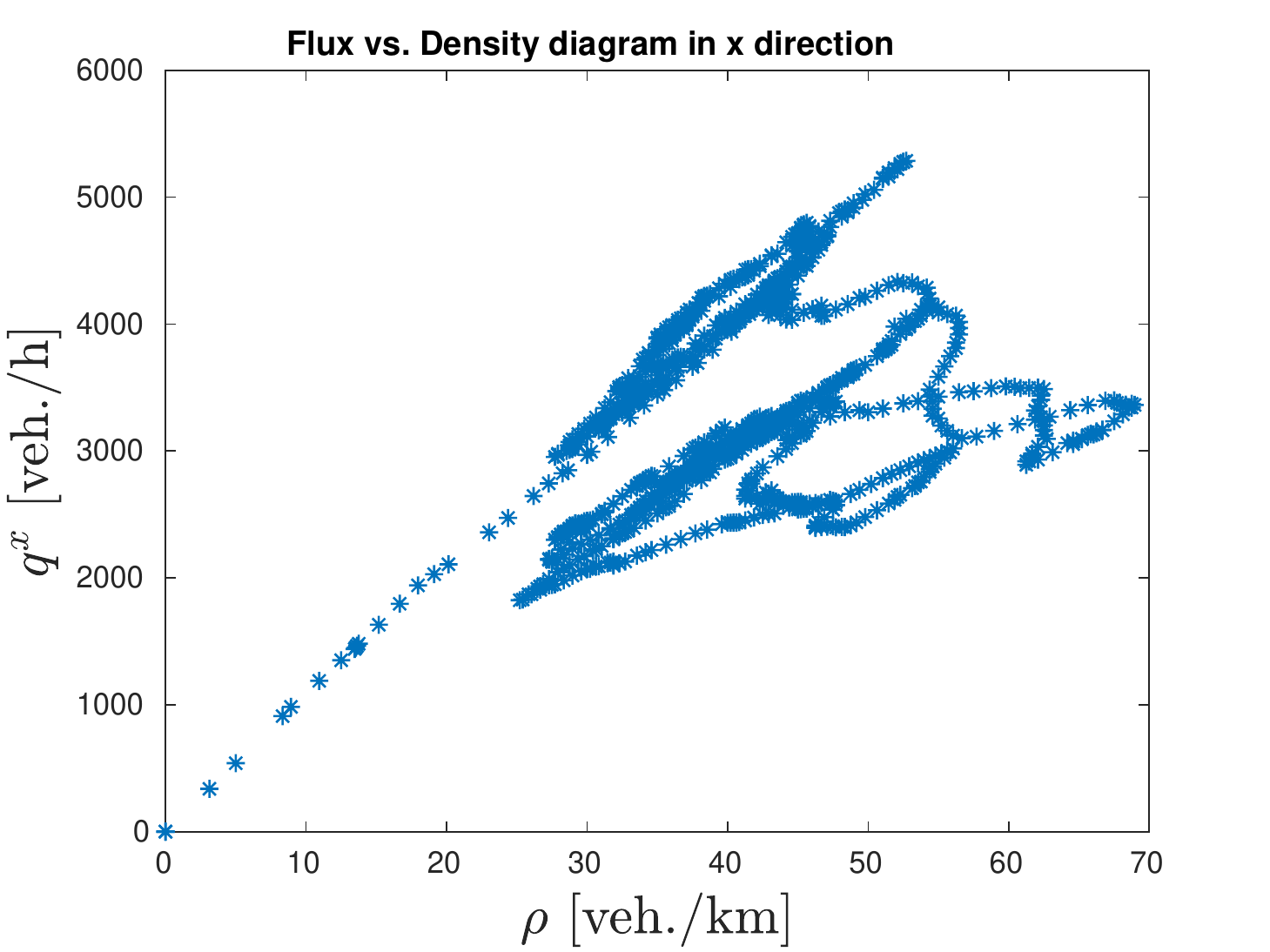}
	\includegraphics[width=0.49\textwidth]{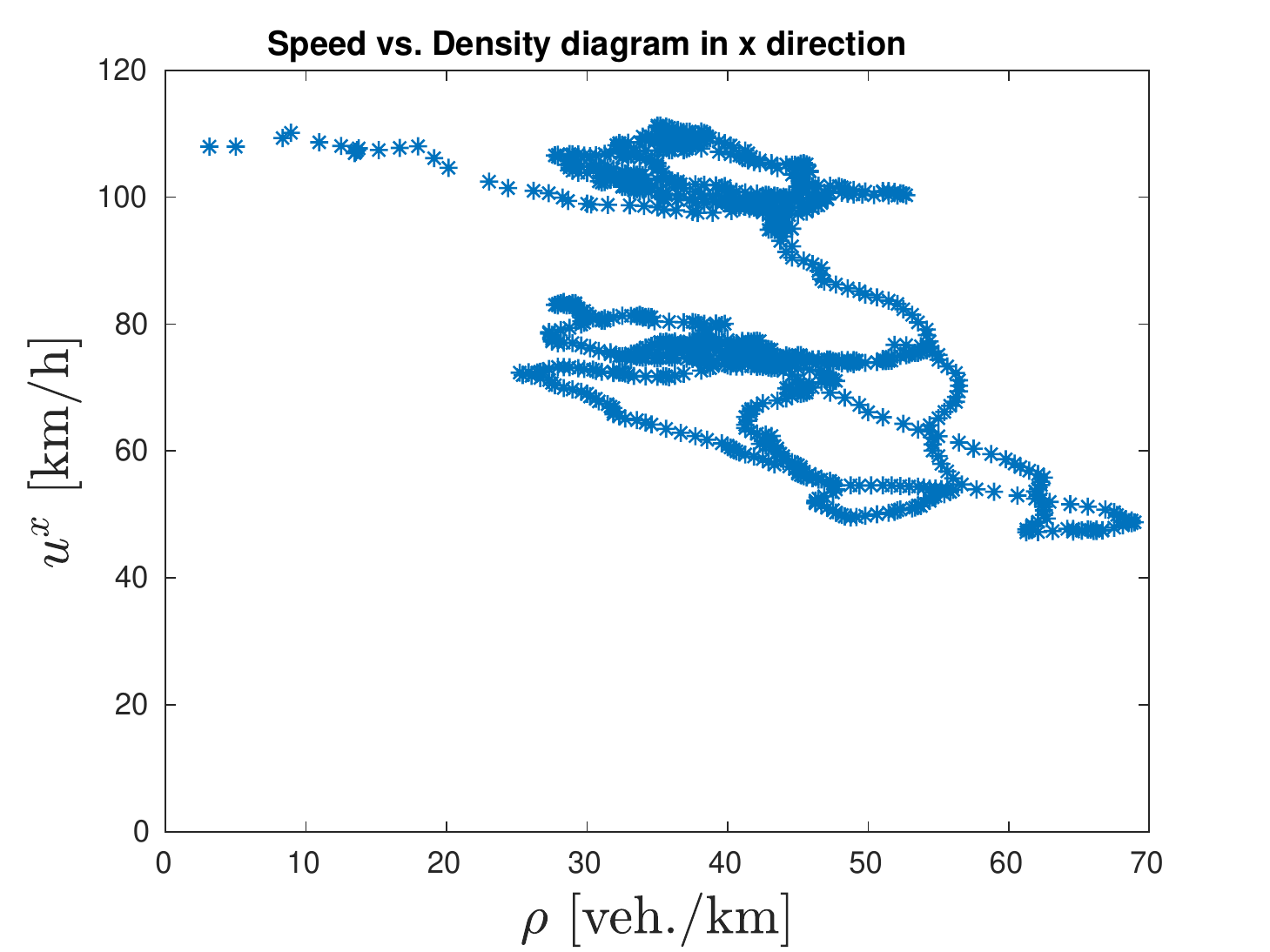}
	\\
	\includegraphics[width=0.49\textwidth]{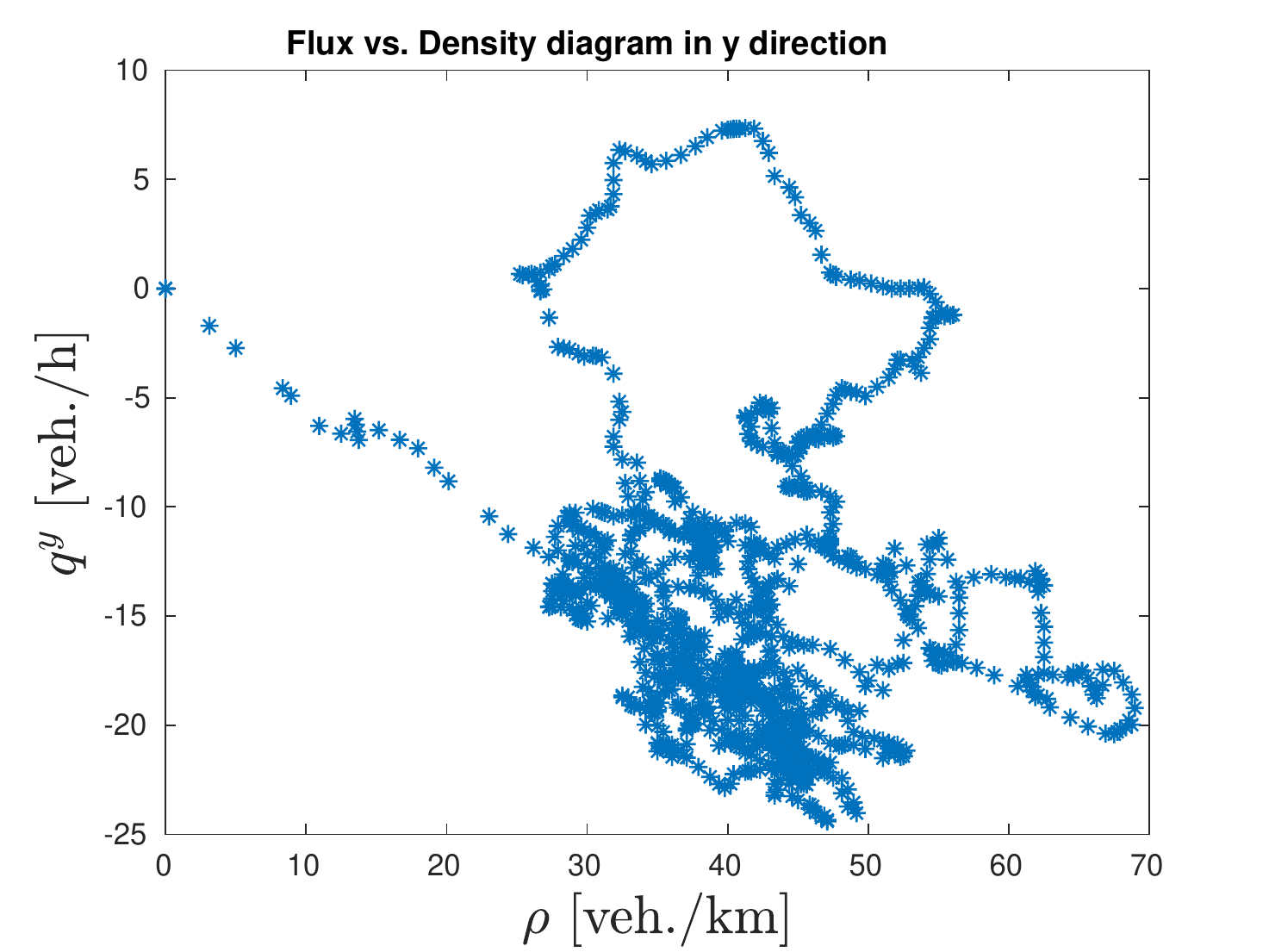}
	\includegraphics[width=0.49\textwidth]{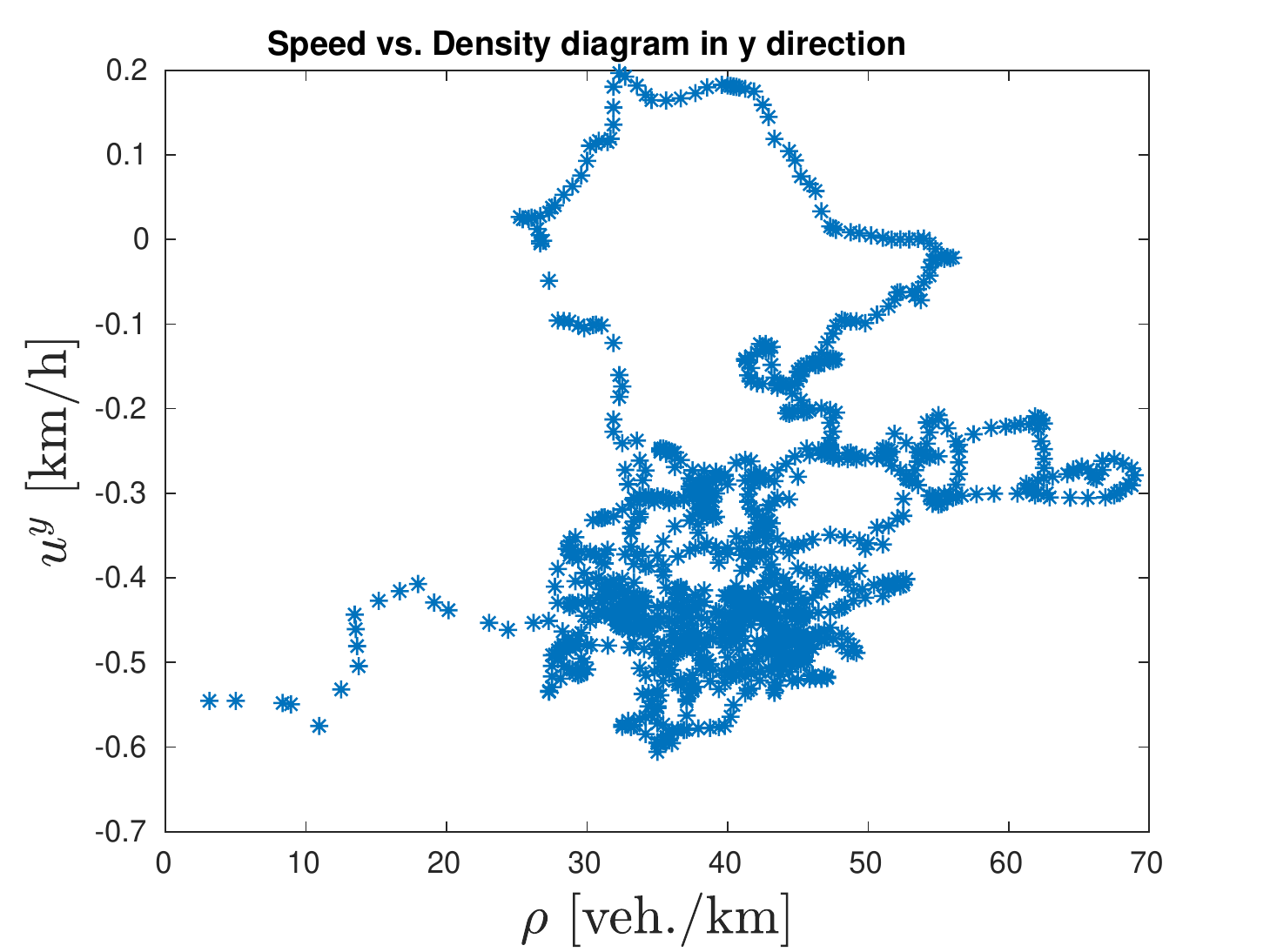}
	\caption{Experimental diagrams from the A3 German highway using $20$~minutes of recorded video. Top row: flux-density (left) and speed-density (right) diagrams in $x$-direction. Bottom row: flux-density (left) and speed-density (right) diagrams in $y$-direction.\label{fig:experimentalFD}}
\end{figure}

In Figure~\ref{fig:experimentalFD} we show the diagrams resulting from the German data-set: the top row shows the relations $(\rho,q^x)$ and $(\rho,u^x)$, while the bottom row shows the relations $(\rho,q^y)$ and $(\rho,u^y)$. Observe that the data-set provides during the time period several levels of congestion but we never observe bumper-to-bumper conditions. In fact, the maximum density is about $70$~vehicles per kilometer.

In addition to the classical fundamental relations $(\rho,q^x)$ and $(\rho,u^x)$ we also show the diagrams $(\rho,q^y)$ and $(\rho,u^y)$. We highlight that data-sets providing 2D trajectories are already available, e.g., see \cite{NGSIM,RTMC}. But the attempt of taking into account the study of the dynamics across the lanes, and thus considering also the data in $y$-direction, is, to our knowledge, a novelty in the mathematical literature on traffic flow.

\begin{figure}[t!]
	\centering
	\vspace{-0.25cm}
	\includegraphics[width=\textwidth]{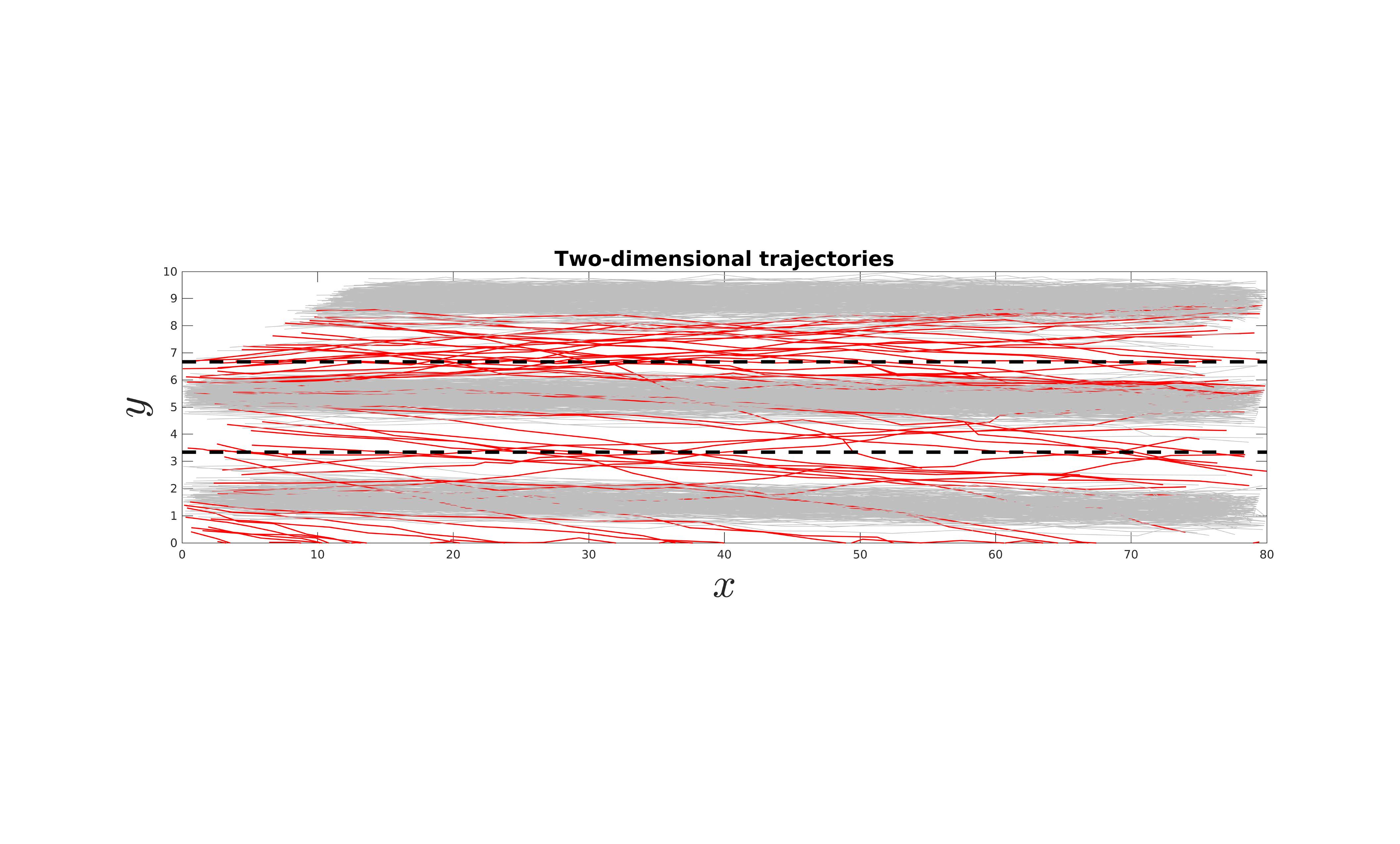}\vspace{-2cm}
	\caption{Two-dimensional trajectories extrapolated from the German data-set. In red we show the trajectories of vehicles crossing a lane while traveling.\label{fig:Trajectories}}
\end{figure}

The qualitative structure of such diagrams is defined by the properties of different regimes, or phases, of traffic. For a description of the diagrams in $x$-direction we refer, e.g., to~\cite{GreenshieldsSymposium,kerner2004BOOK,Leutzbach}. Clearly, the diagrams in $y$-direction show a different quantitative and qualitative behavior with respect to the classical ones. Firstly, we observe that the values of the flux and of the mean speed are about $10^3$ smaller than the values in $x$-direction. This is obvious since the velocity of vehicles along the road is higher then the lateral velocity and thus $u^y$ is not dominant with respect $u^x$. In other words, we are looking at two behaviors occurring at different scales. But this does not mean that the behavior in $y$-direction can be neglected. In fact, an analysis of the trajectories shows that about the $15\%$ of the total vehicles crosses a lane while traveling the road section. See Figure~\ref{fig:Trajectories}. In~\cite{HertyVisconti} we also show, using the same German data-set described above, that the level of potential conflicts is strongly affected by the lane changing. To this end we computed the average risk on the two directions of the flow by means of the Time-To-Collision metric and we observe that the risk due to lane changing is higher.

Moreover, notice that $q^y$ and $u^y$ have positive and negative values since across the lanes vehicles are free to travel in the two directions, towards right and left. Precisely, we assume that positive speeds represent the motion towards the leftmost lane, instead negative speeds represent the motion towards the rightmost lane.

\begin{remark}
	As pointed-out above, the fundamental diagrams in Figure~\ref{fig:experimentalFD} are obtained by averaging over a time period of $T=60$ seconds macroscopic data computed each $1$ second. If we take more frequent time observations (e.g. every $0.2$ seconds) the structure of the diagrams does not change. We only observe that the data become thicker. Compare the top-left panel in Figure~\ref{fig:experimentalFD} with the left panel in Figure~\ref{fig:FDdifferent}. This behavior is due to the fact that we consider a linear approximation of the vehicle trajectories, i.e. we assume that the vehicle velocities are constant. Thus, considering time observations every $1$ second does not keep out information.
	
	In contrast, the time for the data aggregation slightly influence the structure of the fundamental diagrams. Compare the top-left panel in Figure~\ref{fig:experimentalFD} with the right panel in Figure~\ref{fig:FDdifferent}. In the following we will consider the diagrams obtained with an aggregation time period of $T=60$ seconds which is widely used in the engineering literature on traffic flow.
	
	\begin{figure}[t!]
		\centering
		\includegraphics[width=0.49\textwidth]{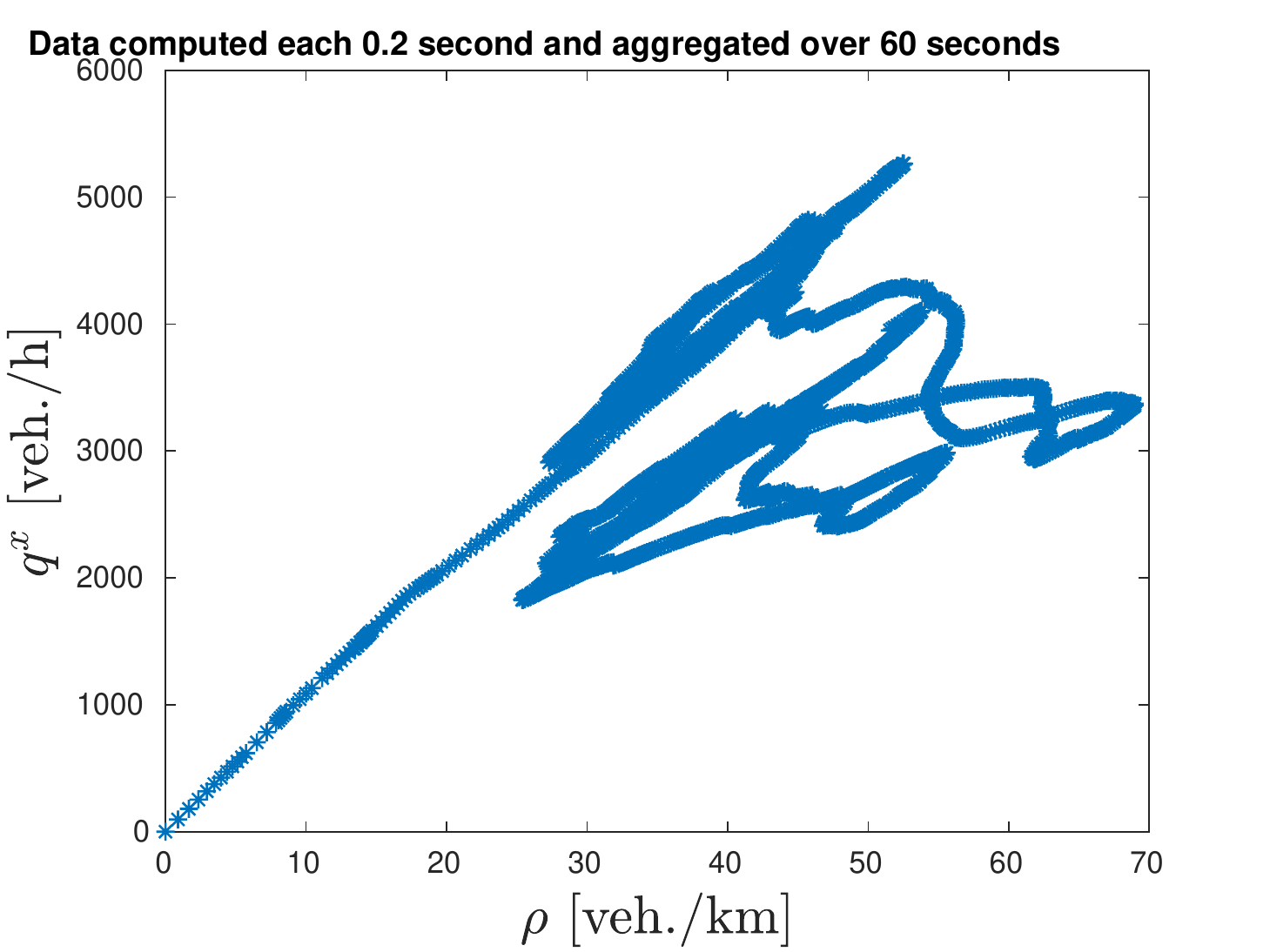}
		\includegraphics[width=0.49\textwidth]{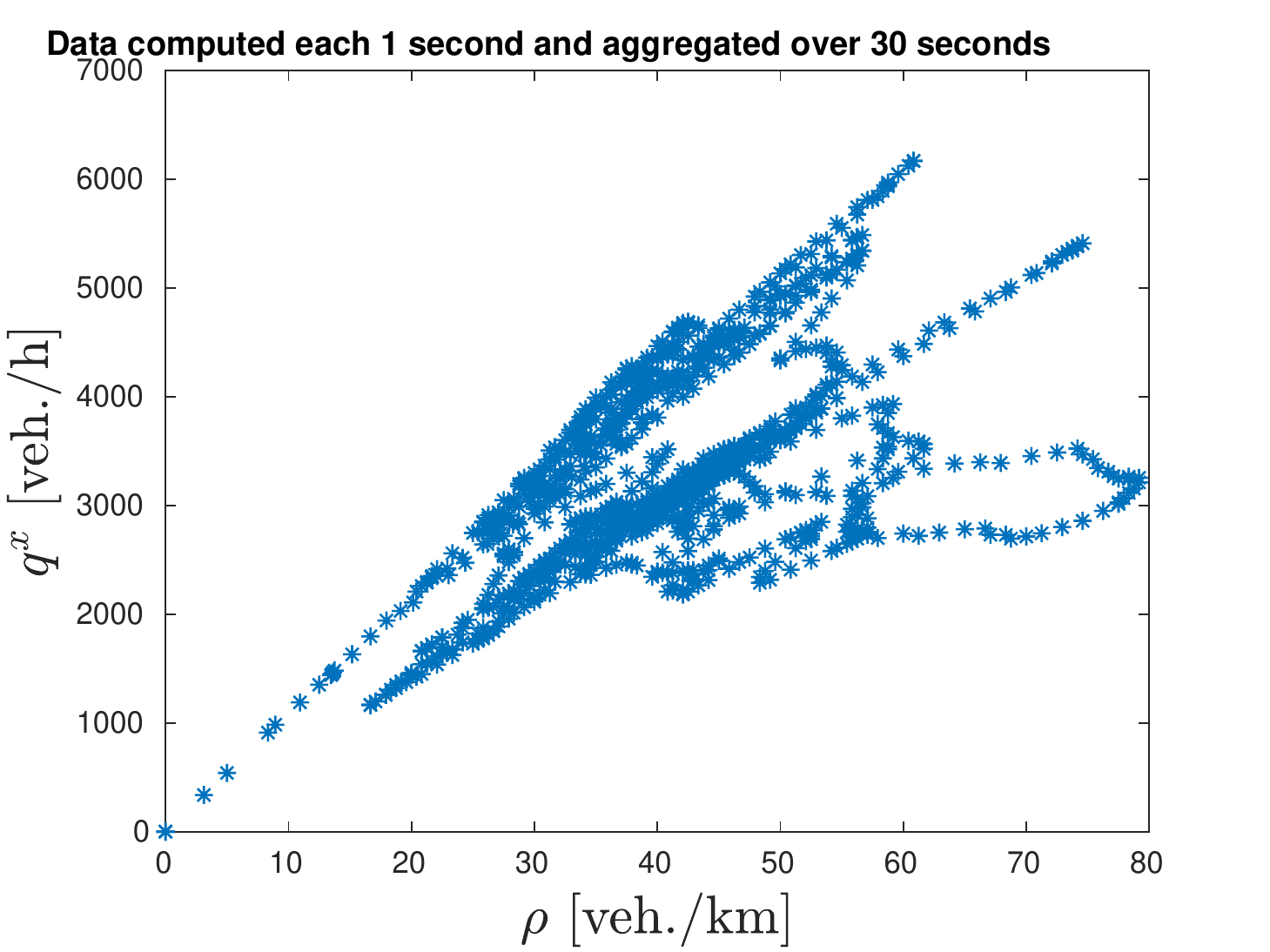}
		\caption{Fundamental diagrams in $x$ direction. Left: macroscopic data computed each $0.2$ seconds and aggregated over $60$ seconds. Right:  macroscopic data computed each $1$ second and aggregated over $30$ seconds.\label{fig:FDdifferent}}
	\end{figure}
\end{remark}

\section{Two-dimensional LWR-type model} \label{sec:MacroModel}

%\subsection{One dimensional models} \label{sec:1DModel}

One dimensional first-order macroscopic traffic models are based on the continuity equation
\begin{equation} \label{eq:continuity_equation}
\partial_t \rho + \partial_x (\rho u) = 0, \quad t\in\mathbb{R}^+,\; x\in[0,L]
\end{equation}
which gives the conservation of vehicles on the road segment $[0,L]$. In \eqref{eq:continuity_equation}, the vehicle density is $\rho(x,t)$, and the vehicle velocity field is $u(x,t)$, where $x$ is the position along the road, and $t$ is time.

The simplest macroscopic traffic model, the LWR  \cite{lighthill1955PRSL,richards1956OR}  model, is obtained by assuming a functional relationship between $\rho$ and $u$, i.e., $u = u(\rho)$. This turns equation \eqref{eq:continuity_equation} into a scalar hyperbolic conservation law
\begin{equation} \label{eq:1DLWR}
\partial_t \rho + \partial_x q(\rho) = 0,
\end{equation}
where the flux $q$ is given by the flow rate function $q(\rho) = \rho u(\rho)$. Because the LWR model \eqref{eq:1DLWR} is a closed model consisting of a single equation, it is denoted a \emph{first-order model}. The velocity function $u(\rho)$ is commonly assumed to be decreasing in $\rho$ with $u(\rho_\text{max}) = 0$ for some maximal vehicle density $\rho_\text{max}>0$. Here, $\rho_\text{max}$ is assumed to be the density in bumper-to-bumper conditions and its value is given in Section~\ref{sec:DataFitting} below.

The strict functional relationship between $\rho$ and $u$ is called \emph{closure law} and is loosened in the so-called \emph{second-order models}, which augment \eqref{eq:continuity_equation} by an evolution equation for the velocity field, see \cite{aw2000SIAP,FanHertySeibold}. 

%\subsection{Two-dimensional LWR-type model} \label{sec:2DModel}

The one-dimensional model \eqref{eq:1DLWR} describes the flow of vehicles in the simple case of a single-lane road or, if the road has multiple lanes (in a given direction), it considers these aggregated into the scalar field quantities $\rho$ and $u$. Nevertheless, the dynamics of traffic on a multi-lane highway could be more complex and is strongly influenced by the motion of vehicles across the lanes. For this reason, we take into account the intrinsic multi-dimensional characteristic of traffic flow by extending model \eqref{eq:1DLWR} to the two-dimensional first-order macroscopic model
\begin{equation} \label{eq:2DLWR}
\partial_t \rho + \partial_x q^x + \partial_y q^y = 0, \quad t\in\mathbb{R}^+,\; x\in[0,L^x],\; y\in[0,L^y]
\end{equation}
where $q^x = \rho u^x$ and $q^y = \rho u^y$ are the fluxes in the two possible directions of the flow and $u^x$, $u^y$ are the speed along the road and the lateral speed, respectively. The quantities $L^x$ and $L^y$ are the length and the width of the road, respectively. Clearly, one expects that $L^y \ll L^x$.  As in the one-dimensional model,  we have the following two closures
\[
q^x(\rho) = \rho u^x(\rho), \quad q^y(\rho) = \rho u^y(\rho).  
\]
The velocity function $u^x(\rho)$ is the same speed $u$ introduced in the one-dimensional LWR model \eqref{eq:1DLWR} and thus it is obvious to assume that it has the same properties discussed previously. A heuristic description of $q^y$ and $u^y$ as function of the density is not immediate since
it depends strongly on the preference on the drivers as well as general imposed traffic rules. It is natural to assume that the lateral speed is $u^y=-V^y_{\max}$ for $\rho\approx0$ and $u^y=0$ $\rho\approx\rho_\text{max}$. In fact, in the first case vehicles would travel towards the right-most lane since the road is free (according to the traffic rules), while in the second case they cannot change lanes and thus travel in $y$-direction since the lanes are almost congested. Note that contrary to the $x$-direction the speed in $y$-direction can be negative. In the following section we propose a functional relation obtained from data. 

We finally stress the fact that the two-dimensional model \eqref{eq:2DLWR} is able to take into account the dynamics of traffic on a multi-lane highway but actually it is not a multi-lane model. In fact, notice that equation \eqref{eq:2DLWR} is continuous in $y$. Instead a multi-lane model requires to treat lanes as discrete object and, thus, to develop a system of balance laws in which the source terms describe the mass exchange between the lanes.

\subsection{Macroscopic closures and data-fitting} \label{sec:DataFitting}

For the dynamics along the road, namely in $x$-direction, several laws have been considered in the literature: popular examples of flow rate functions $q^x(\rho)$ are the Greenshields' flux \cite{Greenshields1935aa}, in which $q^x(\rho)$ is a quadratic function, and the Newell-Daganzo flux \cite{Daganzo1994aa,Newell1993}, in which $u^x(\rho)$ is a piece-wise linear function. These different choices of functions lead to well-posed first-order models. Many closure laws were proposed in the literature, for further discussions we refer, e.g., to the book \cite{Rosini}.

A natural way to derive closure laws  is to construct a fitting of the experimental data. Although this approach ignores the scattered behavior of data, we expect to characterize key properties of the traffic flow  (as the critical density, the maximum flow, $\dots$). For comparisons between models using classical closure laws and data-fitted models, see \cite{FanHertySeibold,FanSeibold2012} based on the NGSIM data-set \cite{NGSIM} and on the RTMC data-set \cite{RTMC}.

We are considering the German data-set described in Section \ref{sec:DataSet} and we are proposing a two-dimensional first-order macroscopic model. Then, in order to get the closure laws to complete equation \eqref{eq:2DLWR} we proceed by constructing the best fitting via a least squares fit to the data computed in Section \ref{sec:DataSet} and showed in Figure \ref{fig:experimentalFD}. The closures for the first-order macroscopic model \eqref{eq:2DLWR} must represent these data via single-valued functions $q^x(\rho)$ and $q^y(\rho)$. 

Since the stagnation density $\rho_\text{max}$ is not represented well via data, we prescribe it as a fixed constant, given by the ratio between the number of lanes and the typical vehicle length of $5$ meters, plus $50\%$ of additional safety distance, so that
\begin{equation*}
\rho_\text{max} = \frac{3\;\text{lanes}}{7.5\;\text{m}} = 400\;\text{veh./km},
\end{equation*}
where this value is obtained by considering the unit distance of 3000 meters for the three lanes (1 kilometer per lane).

As visible in the flux-density diagram in the top left panel of Figure~\ref{fig:experimentalFD}, the data tend to exhibit a relatively linear increasing relationship between $\rho$ and $q^x$ for low densities. In turn, for higher densities, a significant spread is visible, i.e., a single $\rho$ value corresponds to many different flow rates $q^x$. For the data in $x$-direction then we employ the same approach presented in \cite{FanHertySeibold} and in \cite{FanSeibold2012} by selecting a three-parameter family of smooth and strictly concave flow rate curves as
\begin{equation} \label{eq:FitX}
q^x_{\alpha^x,\lambda^x,p^x}(\rho) = \alpha^x\left(d_1+(d_2-d_1)\frac{\rho}{\rho_\text{max}}-\sqrt{1+d_3^2}\right),
\end{equation}
where
\begin{align*}
d_1 = \sqrt{1+\left(\lambda^x p^x\right)^2}\;, \quad
d_2 = \sqrt{1+\left(\lambda^x(1-p^x)\right)^2}\;,\quad
d_3 = \lambda^x \left(\frac{\rho}{\rho_\text{max}}-p^x\right).
\end{align*}
Each flow rate function $q^x_{\alpha^x,\lambda^x,p^x}(\rho)$ in this family vanishes for $\rho = 0$ and $\rho = \rho_\text{max}$. The three free parameters allow for controlling three important features of the flux-density diagram in $x$-direction: the value of maximum flow rate (mainly determined by $\alpha^x$), the critical density (mainly controlled by $p^x$), and the curvature (dominated by $\lambda^x$).

In case of the $y$-direction a significant spread of the data is visible also at lower densities and they show both positive and negative values since the motion is allowed in two directions: towards the left-most and the right-most lane. Moreover, we observe only $u^y<0$ for $\rho\approx 0$, proving the fact that vehicles tend to travel towards the right-most lane in the free-flow regime. To take into account these features we have to propose a different flow rate curve with respect to \eqref{eq:FitX}. Precisely, in this case we choose a simple two-parameter family of smooth functions as follows
\begin{equation} \label{eq:FitY}
q^y_{\alpha^y,p^y}(\rho) = \alpha^y \rho \left(1 - \left( \frac{\rho}{\rho_{\max}} \right)^{p^y} \right).
\end{equation}
Each flow rate function $q^y_{\alpha^y,p^y}(\rho)$ in this family vanishes for $\rho = \rho_\text{max}$. The two free parameters allow for controlling the speed in the free-flow regime (determined by $\alpha^y$) and the shape of the of the curve due to the data (mainly controlled by $p^y$).

\begin{remark}
	Clearly, a more complex flow rate curve \eqref{eq:FitY} may be postulated. The simple choice \eqref{eq:FitY} is justified by the behavior of the $y$-diagrams provided by the A3 German highway which do not give information on how the data behave for higher density values. In fact, the only realistic a-priori assumption for the congested regime is that $u^y = 0$.
\end{remark}

From the three- and the two-parameter family of flow rate curves (equations \eqref{eq:FitX} and \eqref{eq:FitY}, respectively), the closures $q^x$ and $q^y$ are selected in such a way they are the closest, in a least-squares sense, to the experimental data points $(\rho_j,q^x_j)$, and $(\rho_j,q^y_j)$, respectively. Thus we solve
\begin{equation} \label{eq:LSQ}
\min_{\alpha^x,\lambda^x,p^x}\norm{q^x_j-q^x_{\alpha^x,\lambda^x,p^x}(\rho_j)}_2^2, \quad \min_{\alpha^y,p^y}\norm{q^y_j-q^y_{\alpha^y,p^y}(\rho_j)}_2^2.
\end{equation}

\begin{figure}[t!]
	\centering
	\includegraphics[width=0.49\textwidth]{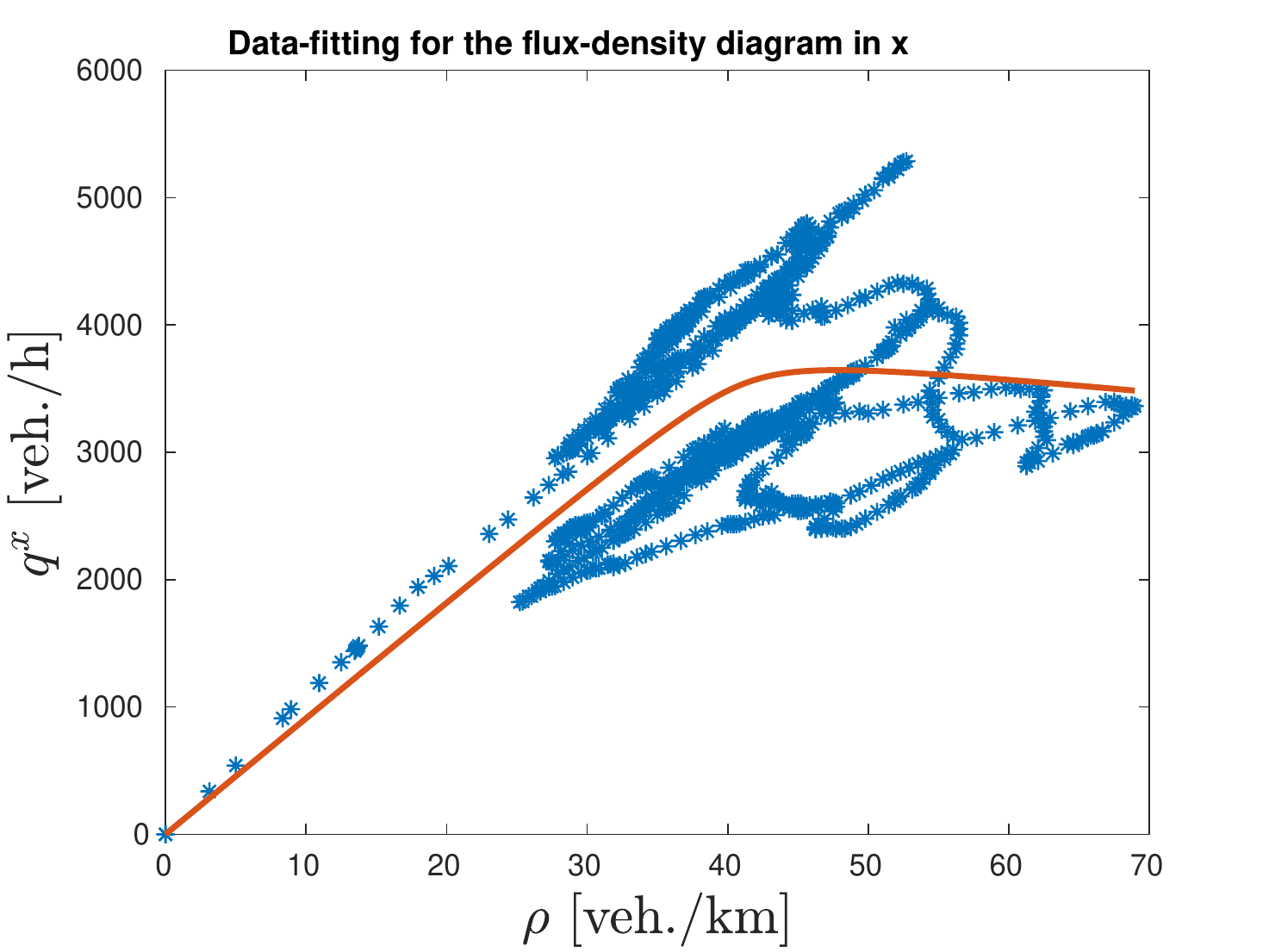}
	\includegraphics[width=0.49\textwidth]{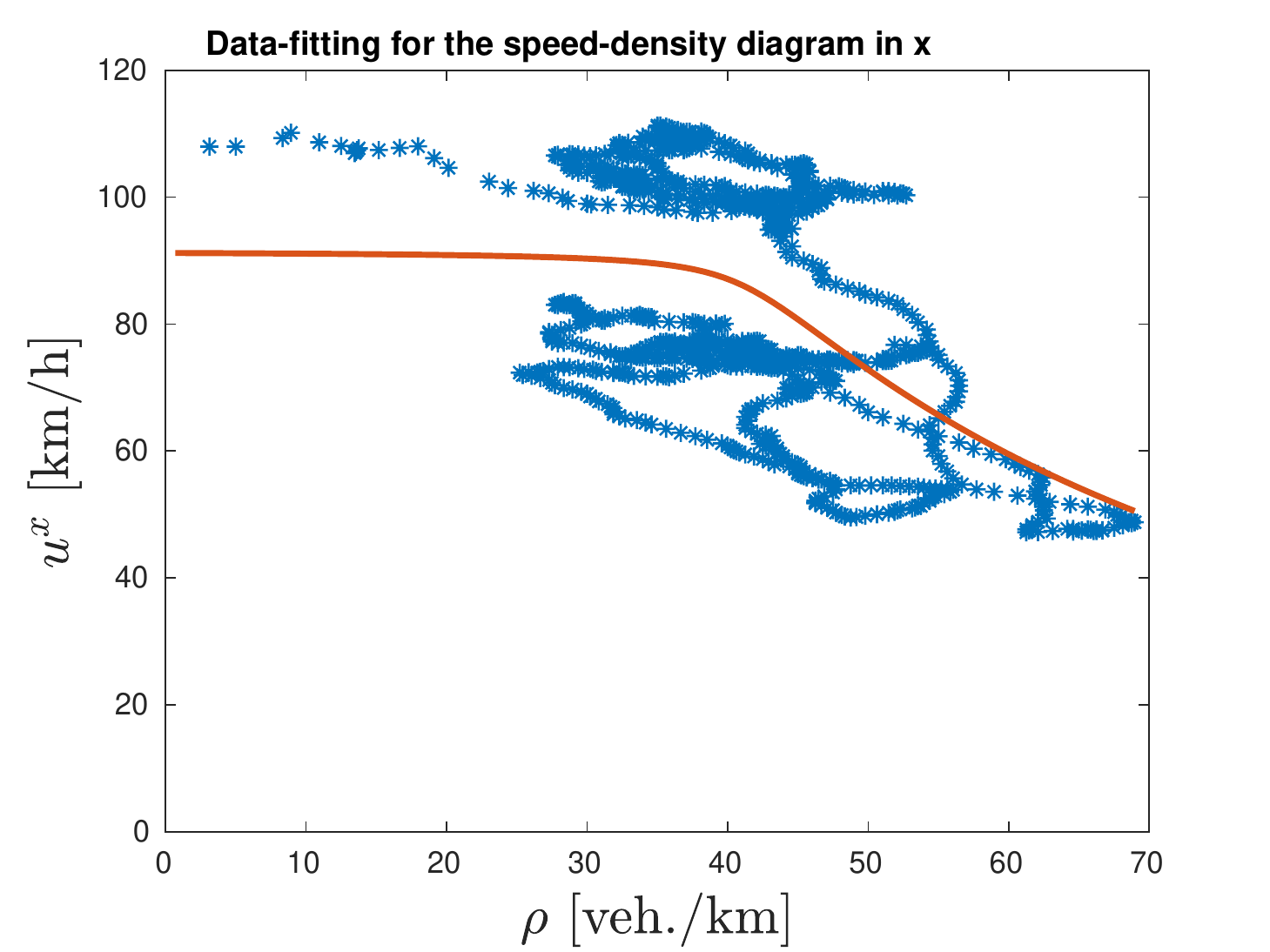}
	\\
	\includegraphics[width=0.49\textwidth]{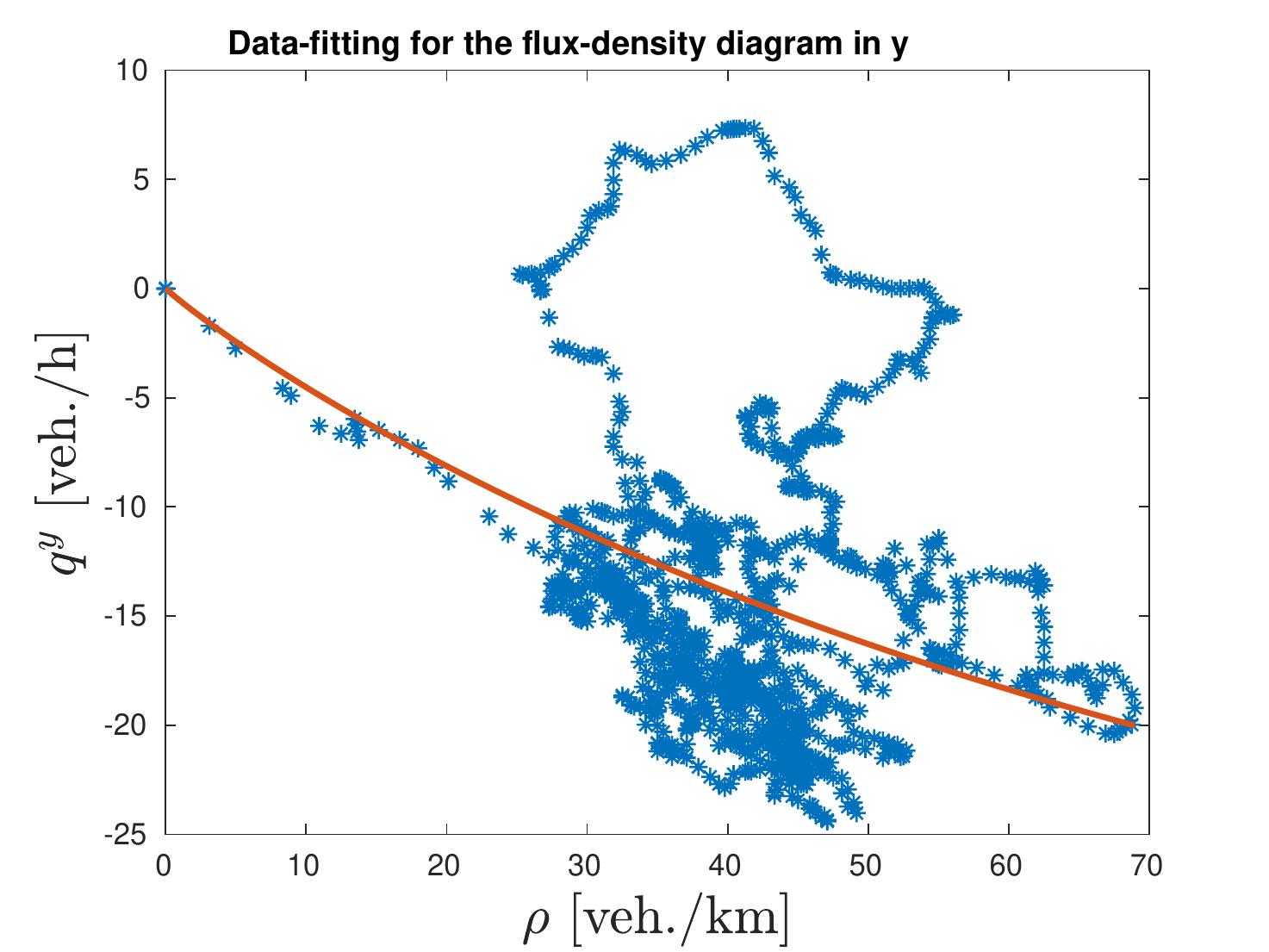}
	\includegraphics[width=0.49\textwidth]{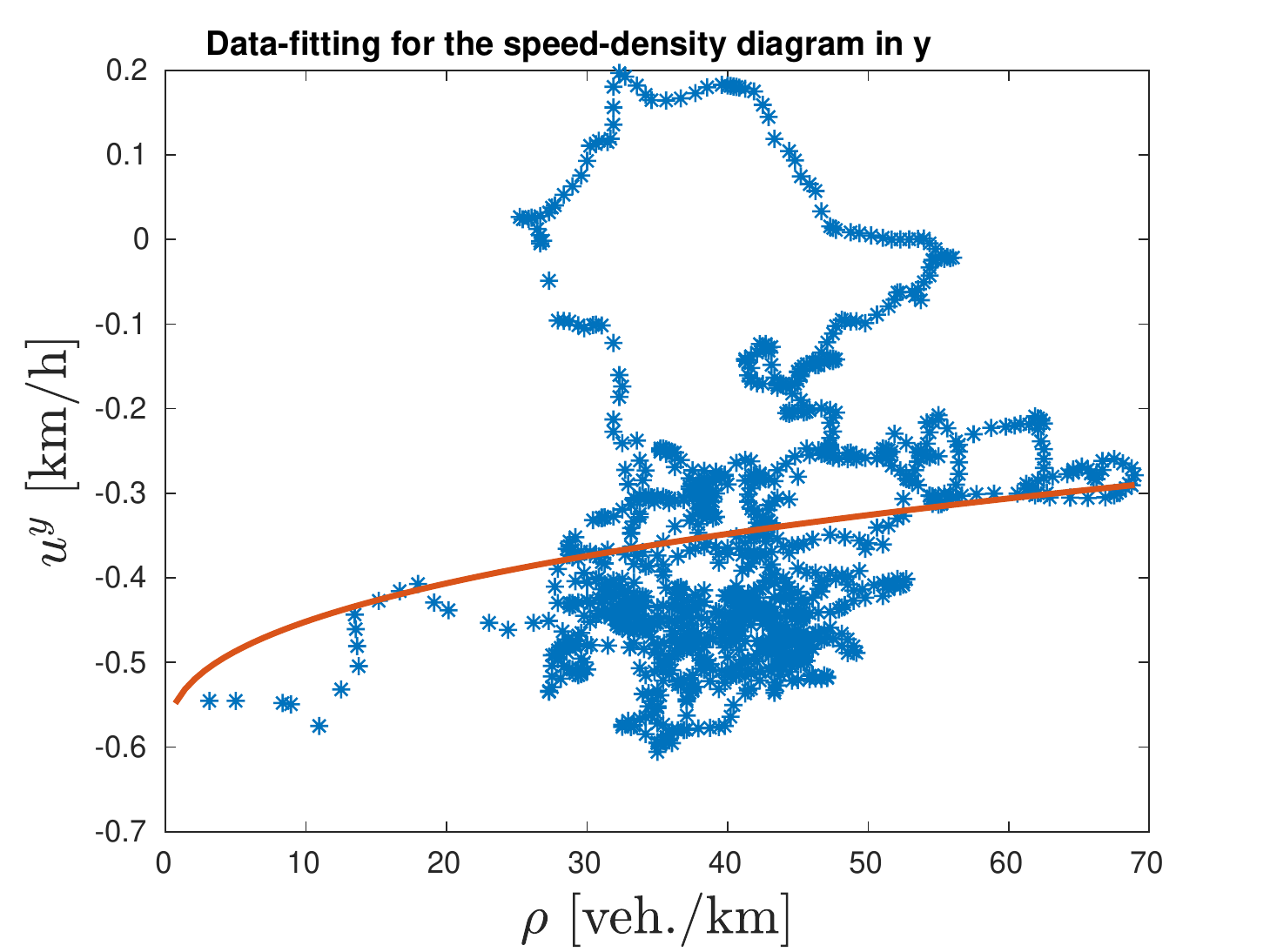}
	\caption{Data-fitting of the experimental diagrams. Top row: functions approximating the flux-density (left) and speed-density (right) diagrams in $x$-direction. Bottom row: functions approximating flux-density (left) and speed-density (right) diagrams in $y$-direction.\label{fig:DataFitted}}
\end{figure}

The minimization problems \eqref{eq:LSQ} are solved numerically by using the Matlab solver \textsf{fmincon} which finds the minimum of constrained nonlinear functions. For the German data-set the solver provides the following values for the free parameters
\begin{itemize}
	\item $\alpha^x = 252.6686$, $\lambda^x = 0.1033$ and $p^x = 80.8620$;
	\item $\alpha^y = -0.6056$ and $p^y = 0.3712$.
\end{itemize}
For the parameters in $x$-directions we do not prescribe a range, i.e. $\left(\alpha^x,\lambda^x,p^x\right)\in\mathbb{R}^3$. While we require that $\alpha^y\in[\min u^y,0]$ and $p^y\in[0,5]$. The constrained minimization problems are quickly solved by the Matlab \textsf{fmincon} function. The CPU time needed is about $0.15$ seconds in each direction. The relative errors obtained with the above optimal parameters are
\[
\frac{\norm{q^x_j-q^x_{\alpha^x,\lambda^x,p^x}(\rho_j)}_2}{\norm{q_j^x}_2} \approx 0.1812, \quad \frac{\norm{q^y_j-q^y_{\alpha^y,p^y}(\rho_j)}_2}{\norm{q_j^y}_2} \approx 0.4.
\]

In Figure~\ref{fig:DataFitted}, the red curves represent the least-squares fits to the given data points computed in Section \ref{sec:DataSet}. These functions are used to close the two-dimensional first-order macroscopic model \eqref{eq:2DLWR} and to validate the model in the next section.

The mathematical properties of the proposed flux in $y$ 
are similar to classical LWR type models. Therefore, a detailed
discussion is skipped. For the numerical scheme below we remark that the conservation law \eqref{eq:2DLWR} with choice \eqref{eq:FitY}  is strictly hyperbolic. Moreover, the optimal parameters $\alpha^y$ and $p^y$ lead to a single inflection point function of the flux function and therefore conservation law \eqref{eq:2DLWR} still give rise to only simple waves (either shock or rarefaction waves) also in $y$-direction.

\section{Numerical simulations} \label{sec:Simulations}

In this section we study the predictive accuracy of the 2D LWR-type model \eqref{eq:2DLWR} with respect to measurement data. In particular, we show that model \eqref{eq:2DLWR} is more accurate than its 1D version \eqref{eq:1DLWR} in which we choose as closure the flow rate function \eqref{eq:FitX}.

To this end, we firstly present the scheme used to numerically solve model \eqref{eq:2DLWR}. Then, we specify how continuous field quantities are constructed from the trajectory data, following the approach described in \cite{FanHertySeibold} and \cite{FanSeibold2012} for 1D data-fitted models.

\subsection{Numerical scheme} \label{sec:Discretization}

In the following, we simply describe the numerical scheme for solving the two-dimensional model \eqref{eq:2DLWR}. For the one-dimensional one \eqref{eq:1DLWR}, the same numerical scheme is used, clearly neglecting the computation of transport term in $y$.

In order to approximate the solution $\rho$ of \eqref{eq:2DLWR}, we use the dimensional splitting method or method of fractional steps, \cite{LevequeBook2002,ToroBook}. We split \eqref{eq:2DLWR} into
\begin{subequations} \label{eq:Split}
	\begin{align}
	\partial_t \rho(t,x,y) + \partial_x q^x(\rho) &= 0, \label{eq:SplitX}\\
	\partial_t \rho(t,x,y) + \partial_y q^y(\rho) &= 0 \label{eq:SplitY}
	\end{align}
\end{subequations}
and for each problem we apply a finite volume approximation. To this end we divide the spatial domain $\Omega = [0,L^x]\times[0,L^y]$ into $N^x\times N^y$ cells $\Omega_{ij} = (x_{i-\nicefrac{1}{2}},x_{i+\nicefrac{1}{2}})\times(y_{j-\nicefrac{1}{2}},y_{j+\nicefrac{1}{2}})$, $i=1,\dots,N^x$, $j=1,\dots,N^y$, such that $\cup_{i,j} \Omega_{ij} = \Omega$ and $x_{i+\nicefrac{1}{2}}-x_{i-\nicefrac{1}{2}}=\Delta x$, $y_{j+\nicefrac{1}{2}}-y_{j-\nicefrac{1}{2}}=\Delta y$. Thus $\abs{\Omega_{ij}}=\Delta x\times\Delta y$.

We consider a semi-discrete finite volume scheme and denote by
\[
\overline{\rho}_{ij}(t) = \frac{1}{\abs{\Omega_{ij}}} \int_{\Omega_{ij}} \rho(t,x,y) \mathrm{d}x \mathrm{d}y
\]
the cell average of the exact solution in the cell $\Omega_{ij}$ at time $t$ and $\overline{U}_{ij}(t)$ its numerical approximation. By integrating each equation \eqref{eq:Split} over $\Omega_{ij}$, dividing by $\abs{\Omega_{ij}}$, using the midpoint rule and finally a $s$-stage Strong Stability Preserving Runge-Kutta method (SSPRK) with Butcher's tableau $(A,b)$ and time step $\Delta t$, we get the fully discrete scheme
\begin{subequations}
	\begin{align}
	\overline{U}_{ij}^* &= \overline{U}_{ij}^n - \frac{\Delta t}{\Delta x} \sum_{k=1}^s b_i \left( F_{i+\nicefrac{1}{2},j}^{(k)} - F_{i-\nicefrac{1}{2},j}^{(k)} \right), \quad i=1,\dots,N^x \label{eq:split1}\\
	\overline{U}_{ij}^{n+1} &= \overline{U}_{ij}^* - \frac{\Delta t}{\Delta y} \sum_{k=1}^s b_i \left( G_{i,j+\nicefrac{1}{2}}^{*(k)} - G_{i,j-\nicefrac{1}{2}}^{*(k)} \right), \quad j=1,\dots,N^y. \label{eq:split2}
	\end{align}
\end{subequations}
giving the approximation of the solutions at time $t^{n+1} = T_{\text{init}} + (n+1) \Delta t$, where $T_{\text{init}}$ is the initial time. Notice that in the $x$-sweeps we start with the cell averages $\overline{U}_{ij}^n$ at time $t^n$ and solve $N^y$ one-dimensional problems with $j$ fixed updating $\overline{U}_{ij}^n$ to $\overline{U}_{ij}^*$. In the $y$-sweeps we then use the $\overline{U}_{ij}^*$ values as data for solving the $N^x$ problems with $i$ fixed, which results in $\overline{U}_{ij}^{n+1}$. Here,
\[
F_{i+\nicefrac{1}{2},j}^{(k)} = \mathcal{F}\left(\overline{U}_{i+\nicefrac{1}{2},j}^{(k),+},\overline{U}_{i+\nicefrac{1}{2},j}^{(k),-}\right), \quad G_{i,j+\nicefrac{1}{2}}^{*(k)} = \mathcal{G}\left(\overline{U}_{i,j+\nicefrac{1}{2}}^{*(k),+},\overline{U}_{i,j+\nicefrac{1}{2}}^{*(k),-}\right),
\]
for $k=1,\dots,s$, are the numerical fluxes approximating $q^x(\rho(t,x_{i+\nicefrac{1}{2}},y_j))$ and $q^y(\rho(t,x_{i},y_{j+\nicefrac{1}{2}}))$, respectively. We consider  $\mathcal{F}$ and $\mathcal{G}$  as local Lax-Friedrichs fluxes. We could also use the Godunov scheme which is less diffusive. However, we choose Lax-Friedrichs for its ease. Instead, $\overline{U}_{i+\nicefrac{1}{2},j}^{(k),\pm}$ and $\overline{U}_{i,j+\nicefrac{1}{2}}^{*(k),\pm}$ are the reconstructions at the cell interfaces, at left and right sides, from the stage values $\overline{U}_{i,j}^{(k)}$ and $\overline{U}_{i,j}^{*(k)}$, respectively. The stage values of the cell averages are evolved by 
\begin{align*}
\overline{U}_{ij}^{(k)} &= \overline{U}_{ij}^n - \frac{\Delta t}{\Delta x} \sum_{\ell=1}^{k-1} a_{k\ell} \left( F_{i+\nicefrac{1}{2},j}^{(\ell)} - F_{i-\nicefrac{1}{2},j}^{(\ell)} \right), \quad k=1,\dots,s\\
\overline{U}_{ij}^{*(k)} &= \overline{U}_{ij}^* - \frac{\Delta t}{\Delta y} \sum_{\ell=1}^{k-1} a_{k\ell} \left( G_{i,j+\nicefrac{1}{2}}^{*(\ell)} - G_{i,j-\nicefrac{1}{2}}^{*(\ell)} \right),  \quad k=1,\dots,s,
\end{align*}
where the $k$-th stage value is assumed to be at time $t^n+c_k \Delta t$.

Without using additional tools, the scheme described above is first-order accurate. In order to get a second-order scheme the following ingredients are necessary. The reconstruction at the interfaces from the stage values is performed using a piece-wise linear reconstruction in each direction. To guarantee the non-oscillatory nature of the reconstruction, we apply a nonlinear limiter for the computation of the slopes and here we use the minmod slope limiter. Thus, e.g.,
\[
\overline{U}_{i+\nicefrac{1}{2},j}^{(k),-} = \overline{U}_{i,j}^{(k)} + \frac{\Delta x}{2} \sigma_i, \quad \overline{U}_{i,j+\nicefrac{1}{2}}^{*(k),-} = \overline{U}_{i,j}^{*(k)} + \frac{\Delta y}{2} \sigma^*_j
\]
where
\begin{align*}
\sigma_i &= \text{minmod}\left( \frac{\overline{U}_{i,j}^{(k)} - \overline{U}_{i-1,j}^{(k)}}{\Delta x}, \frac{\overline{U}_{i+1,j}^{(k)} - \overline{U}_{i,j}^{(k)}}{\Delta x} \right), \\
\sigma^*_j &= \text{minmod}\left( \frac{\overline{U}_{i,j}^{*(k)} - \overline{U}_{i,j-1}^{*(k)}}{\Delta y}, \frac{\overline{U}_{i,j+1}^{*(k)} - \overline{U}_{i,j}^{*(k)}}{\Delta y} \right)
\end{align*}
and the minmod function is defined as
\[
\text{minmod}(a,b) = \begin{cases}
a, \quad \abs{a}<\abs{b} \ \text{and} \ ab>0\\
b, \quad \abs{a}>\abs{b} \ \text{and} \ ab>0\\
0, \quad ab<0
\end{cases}.
\]
For further details we refer, e.g., to \cite{Harten1983,VanLeer1977}.

The dimensional splitting \eqref{eq:split1}-\eqref{eq:split2} is only first-order accurate. See~\cite{Godunov}. For a second-order scheme the Strang splitting technique~\cite{Strang} has to be employed. This method consists in a slight different application of the equations \eqref{eq:split1}-\eqref{eq:split2}. More precisely, equation~\eqref{eq:split1} is used to obtain the update up to time $t^n+\frac{\Delta t}{2}$, i.e. with time step $\frac{\Delta t}{2}$. This datum is then used in equation~\eqref{eq:split2}. Finally, the datum resulting from~\eqref{eq:split2} is used to compute the approximation of the solution at time $t^{n+1}$ by means of equation~\eqref{eq:split1} starting from the time level $t^n+\frac{\Delta t}{2}$, thus with time step $\frac{\Delta t}{2}$. For further details we refer to~\cite{LevequeBook2002,Strang}.

A time-stepping of (at least) second-order is mandatory for all subproblems described in the Strang splitting. Here, as Runge-Kutta scheme we take the Heun's method \cite{Heun} whose coefficients $\{a_{ij}\}_{i,j=1}^s$, $\{b_i\}_{i=1}^s$ and $\{c_k\}_{k=1}^s$ are defined in the following Butcher tableau
\begin{equation*}	
\begin{array}{c|cc}
c_1 & a_{11} & a_{12} \\
c_2 & a_{21} & a_{22} \\
\hline
&&\\[-2ex]
& b_1 & b_2
\end{array}
\quad = \quad
\begin{array}{c|cc}
0 & 0 & 0 \\
1 & 1 & 0 \\
\hline
&&\\[-2ex]
& \tfrac12 & \tfrac12
\end{array} \ .
\end{equation*}
The time step $\Delta t$ is chosen in such a way it satisfies the CFL condition \cite{CourantFriedrichsLewy1928aa}. In particular, if not explicitly specified, in the following we will consider as CFL $0.45$ and the time step $\Delta t$ is
\[
\Delta t = 0.45 \min\left\{ \frac{\Delta x}{\max (q^x)^\prime(\rho)} , \frac{\Delta y}{\max (q^y)^\prime(\rho)} \right\},
\]
where the maximum of the derivative of the flux functions is computed on the density profile $\rho$ at initial time.

For the numerical solution of the one-dimensional LWR model \eqref{eq:1DLWR} we consider the natural one-dimensional version of the second-order finite volume scheme presented above.

The scheme described above is a second-order scheme and for our purposes is sufficient. We choose to employ a dimensional splitting technique since it is conceptually easy to understand, allowing to take advantage of using classical one-dimensional methods for conservation laws. There are also methods for multidimensional conservation laws that are intrinsically multidimensional, see e.g.~\cite{Natalini}. These methods should be used to get more accurate numerical schemes and in this case high-order spatial reconstructions~\cite{CiPgSmVg,Shu:2009:WENOreview} combined with high-order Runge-Kutta schemes have to be considered.

\begin{figure}[t!]
	\centering
	\includegraphics[width=0.49\textwidth]{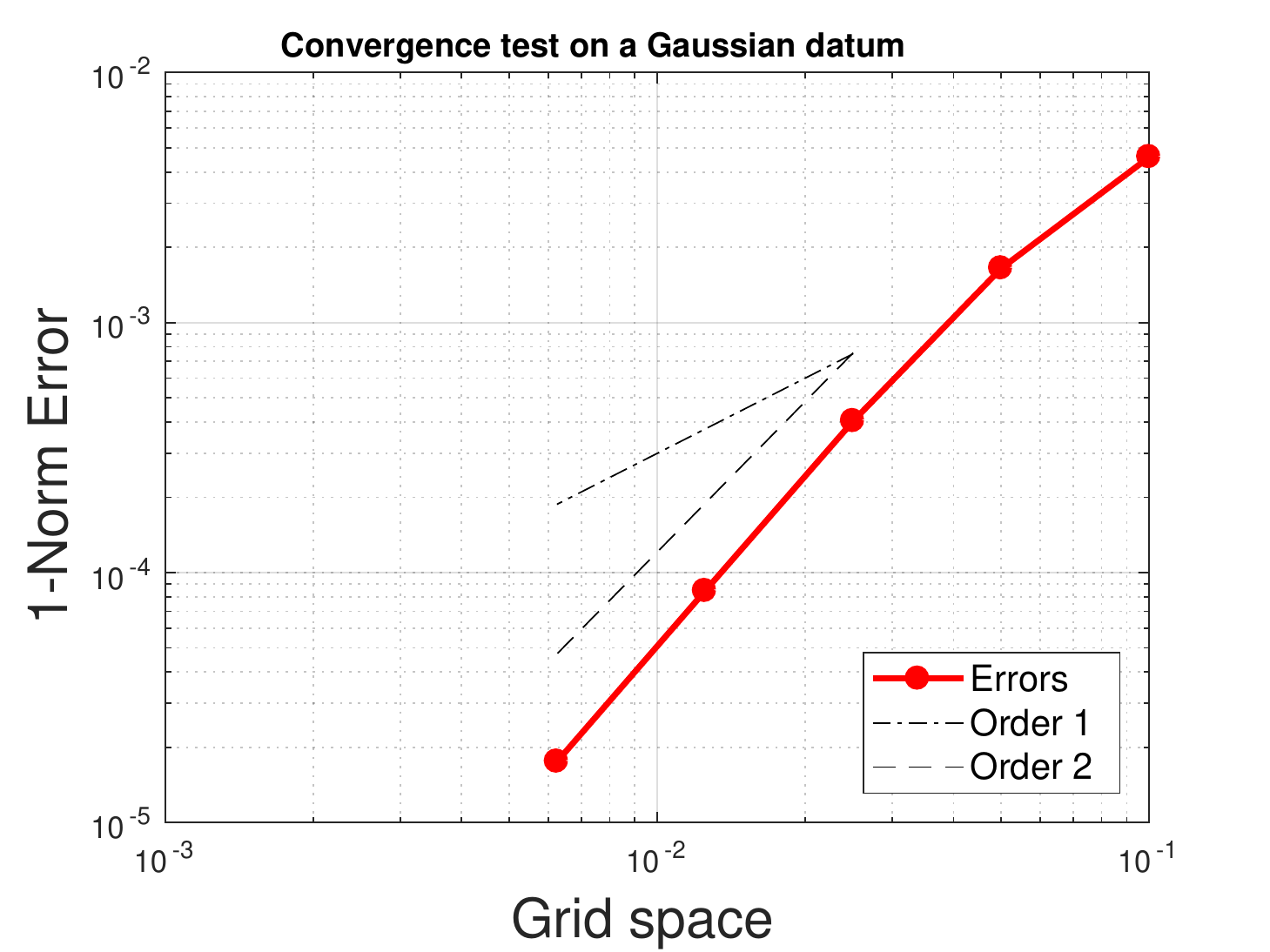}
	\includegraphics[width=0.49\textwidth]{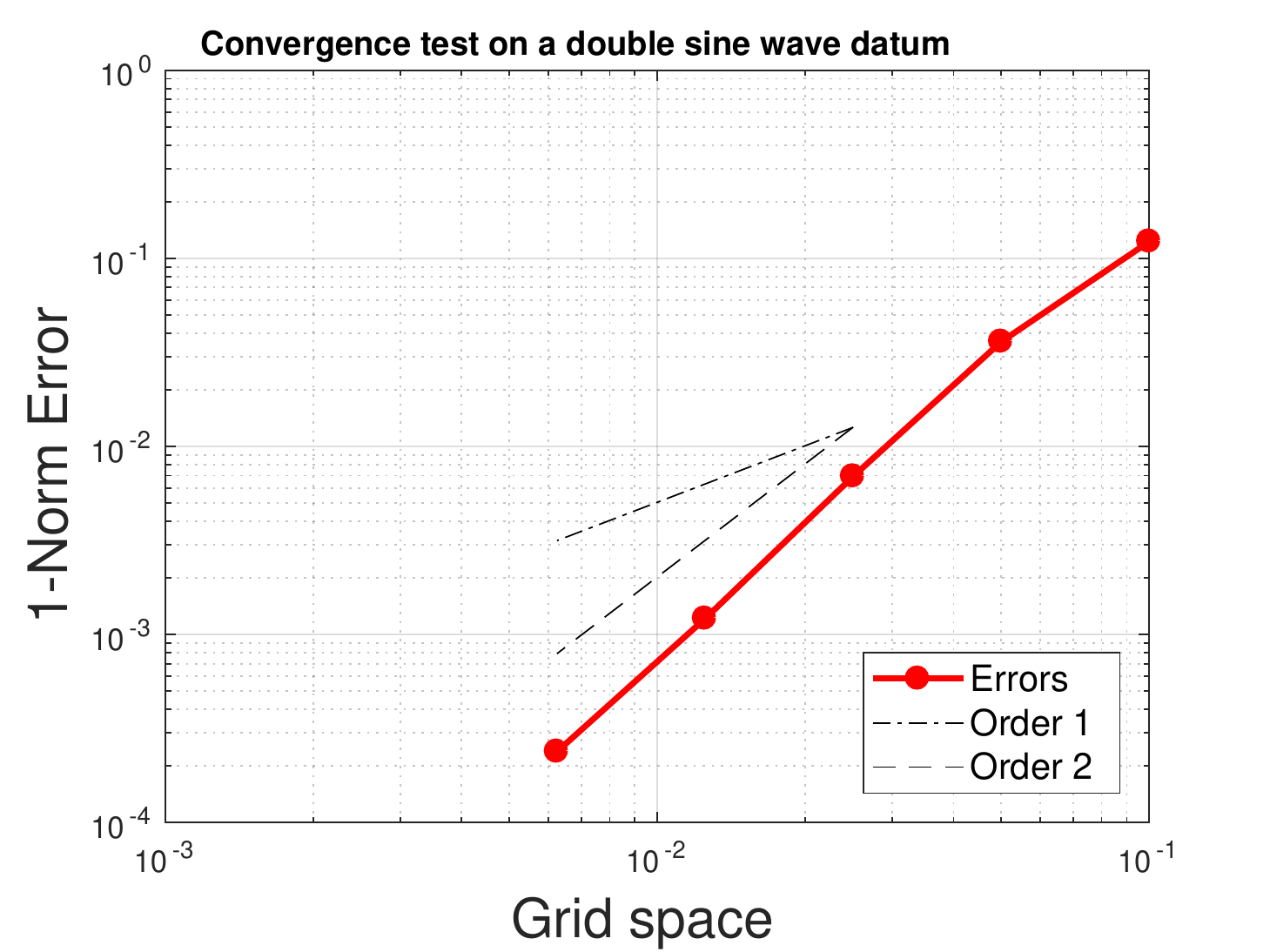}
	\caption{Log-log graph of the convergence test on the Gaussian initial datum (left panel) and on the double sine wave (right panel).\label{fig:convergencetest}}
\end{figure}

Although the scheme presented here is already well-known, for the sake of completeness in Figure~\ref{fig:convergencetest} we show the order of convergence for the case of the two-dimensional linear scalar conservation law
$$
\partial_t \rho(t,x,y) + \partial_x \rho(t,x,y) + \partial_y \rho(t,x,y) = 0, \quad (x,y) \in [-1,1]\times[-1,1]
$$
with a Gaussian initial datum
$$
\rho_0(x,y) = \frac15 e^{-30(x^2+y^2)}
$$
up to time $T_{\text{fin}} = 2$ (left panel of Figure~\ref{fig:convergencetest}) and with a double sine wave initial datum
$$
\rho_0(x,y) = \sin(2\pi x)\sin(2\pi y)
$$
up to time $T_{\text{fin}} = 2$ (right panel of Figure~\ref{fig:convergencetest}). In both cases we use periodic boundary conditions. It is well-known that the scheme still remains second-order accurate also for nonlinear conservation laws on smooth solutions.

\subsection{Treatment of experimental data} \label{sec:DataTreatment}

The numerical implementation of the macroscopic traffic models \eqref{eq:1DLWR} and \eqref{eq:2DLWR}, require the knowledge of continuously in space initial data. Since we are aimed to compare the predictive accuracy of the two models against the data-set described in Section \ref{sec:DataSet}, here we specify how continuous field quantities can be constructed from trajectory data. In particular, we follow the same approach used in \cite{FanHertySeibold} and \cite{FanSeibold2012}. The same idea is applied for computing the reference data in order to compare the
model predictions.

From the German data-set we have the two-dimensional trajectories of vehicles $(x_i(t),y_i(t))$, with a temporal resolution 0.2 seconds, that is essentially continuous in time. However, at each time, the vehicle positions are discrete. In order to incorporate this data as initial condition into the continuous models \eqref{eq:1DLWR} and \eqref{eq:2DLWR}, we must generate a function $\rho(t,x,y)$, for $t=T_{\text{init}}$, that is defined everywhere on the road segment. This approach also allows to compare the model accuracy against the experimental data, constructing at a certain final time the continuous function $\rho(t,x,y)$ from the discrete positions of vehicles with $t=T_{\text{fin}}$.

The construction of density functions from discrete samples is a statistic problem. We employ a kernel density estimation (KDE) approach, with a fixed Gaussian kernel. The specific KDE approach used here is described in \cite{FanSeibold2012} for the case of traffic models and it is called the Parzen-Rosenblatt window method \cite{Parzen1962,Rosenblatt1956}: Assume that at time $t$ we have the positions of vehicles on the road. This data are interpreted as a finite sample of some (unknown) density function. The goal is to reconstruct a kernel density estimator from the discrete information that is close to. At each time instant $t$, KDE starts with a two-dimensional comb function
\[
C(x,y) = \sum_{i=1}^{N(t)} \delta\big(x-x_i(t),y-y_i(t)\big)
\]
where $\delta\big(x-x_i(t),y-y_i(t)\big):=\delta\big(x-x_i(t)\big)\otimes\delta\big(y-y_i(t)\big)$ is the two-dimensional Dirac delta function and $N(t)$ the number of vehicles on the road section at time $t$. Thus the function $C$ accounts for the positions of vehicles on the road at time $t$. Clearly, $C$ cannot be used as initial condition of numerical simulations but we need to define its smoothed version. To this end, we consider a two-dimensional Gaussian kernel
\begin{equation} \label{eq:K}
K(x,y) = \frac{1}{2\pi h^x h^y} e^{-\frac{1}{2}\left(\frac{x}{h^x}\right)^2-\frac{1}{2}\left(\frac{y}{h^y}\right)^2}
\end{equation}
and we define the density function at time $t$ as
\begin{equation} \label{eq:DensityEstimation}
\rho(t,x,y) = \int_{\Omega} K(x-\xi,y-\eta) C(\xi,\eta) \mathrm{d}\xi \mathrm{d}\eta = \sum_{i=1}^{N(t)} K\big(x-x_i(t),y-y_i(t)\big).
\end{equation}
Here $h^x$ and $h^y$ are the bandwidths in $x$- and $y$-direction respectively. There are several works dealing with the optimal choice of the bandwidth in the KDE approach, e.g., see \cite{CaoCuevasManteiga1994,JonesMarronSheather1996}. Here we take the same values already used in \cite{FanHertySeibold,FanSeibold2012} in which the bandwidths are chosen in such a way equally distant vehicles lead to an almost constant density profile. It is clear that this technique for choosing the bandwidths does not depend on the type of data but only on the road section dimensions. Therefore, we recompute the values given in~\cite{FanHertySeibold,FanSeibold2012} for the case of a $80\times 12$ meter road and then we get the kernel widths $h^x = 4$ meter and $h^y = 0.6$ meter.

Finally, we note that the road section is $80$ meters and therefore vehicles travel from the initial point to the exiting one in about $2.7$ seconds, at the maximum speed. 

\subsection{Validation of the model} \label{sec:Validation}

In the following, we validate the presented two-dimensional first-order macroscopic model \eqref{eq:2DLWR} by comparing the evolution of the model with the corresponding measured trajectories. Also, we compare the predictive accuracy of the model with respect to its one-dimensional version \eqref{eq:1DLWR}. 

The deviation between predicted and real traffic states quantifies the model error. Thus we choose the spatial discretization sufficiently fine, namely $\Delta x = \Delta y = 0.5$ meter.

In order to quantify the deviation of the model predictions from the real data, we proceed as follows. Firstly, we compute the continuous density that defines the starting condition at a fixed initial time $T_{\text{init}}$ as in \eqref{eq:DensityEstimation}. Then, we numerically evolve the density profile up to a final time $T_{\text{fin}} > T_{\text{init}}$ using the numerical scheme defined in Section \ref{sec:Discretization} and applied to the model \eqref{eq:2DLWR}. The numerical simulation gives the data output $\mathbf{\overline{U}}\left(T_{\text{fin}}\right)$. The continuous reference solution at time $T_{\text{fin}}$ is constructed from the real data by means of the density estimation defined in \eqref{eq:DensityEstimation}. From this function we obtain  discrete values $\mathbf{\overline{U}}^{\text{exact}}\left(T_{\text{fin}}\right)$,  and we finally compute the prediction error as 
\begin{equation} \label{eq:Error}
E\left( T_{\text{fin}} \right) = \norm{\mathbf{\overline{U}}\left(T_{\text{fin}}\right)-\mathbf{\overline{U}}^{\text{exact}}\left(T_{\text{fin}}\right)}_{L^1(\mathbb{R})}.
\end{equation}

\subsubsection{Predictive accuracy against trajectory data}

\begin{figure}[t!]
	\centering
	\includegraphics[width=0.49\textwidth]{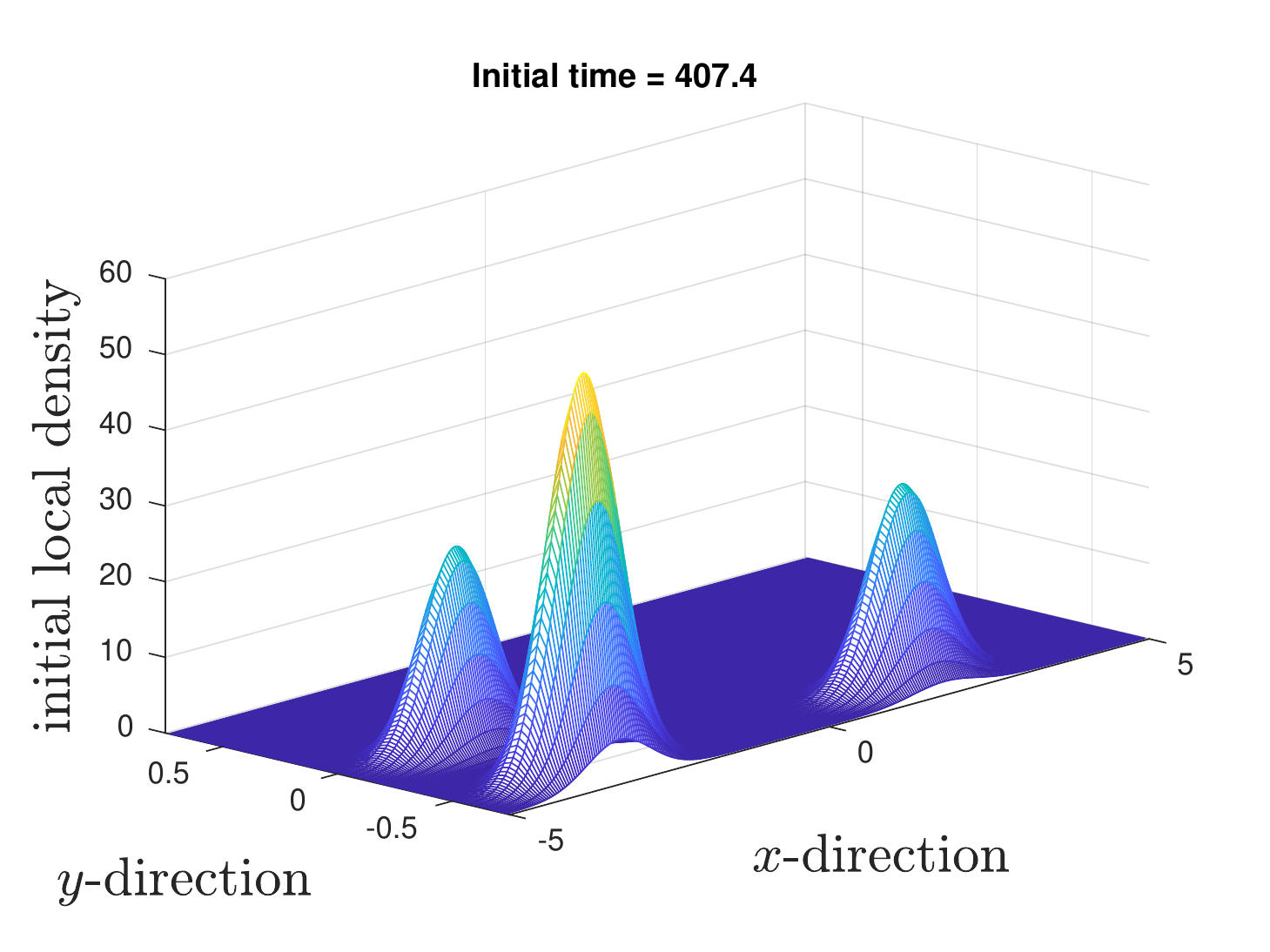}
	\includegraphics[width=0.49\textwidth]{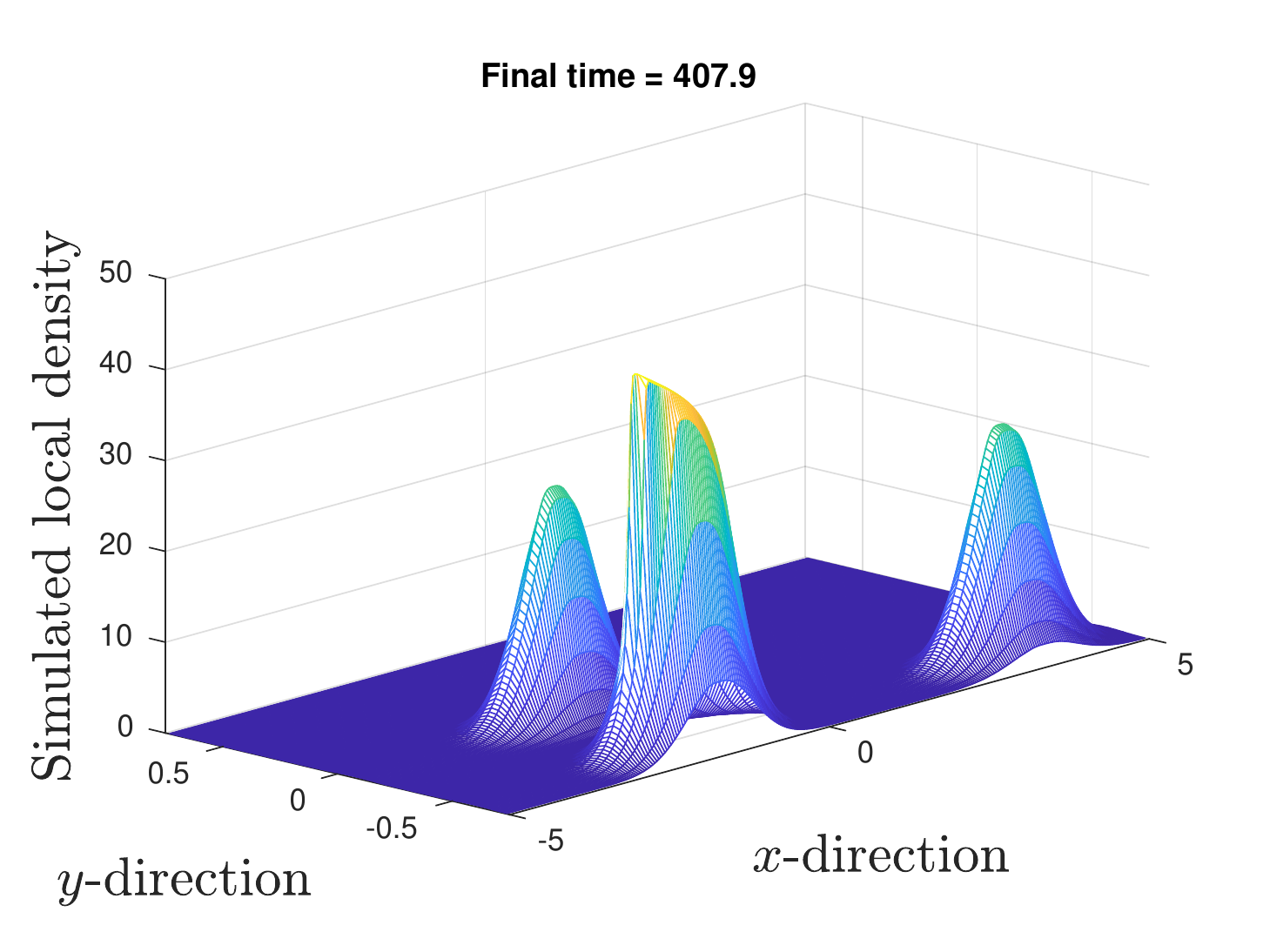}\\
	\includegraphics[width=0.49\textwidth]{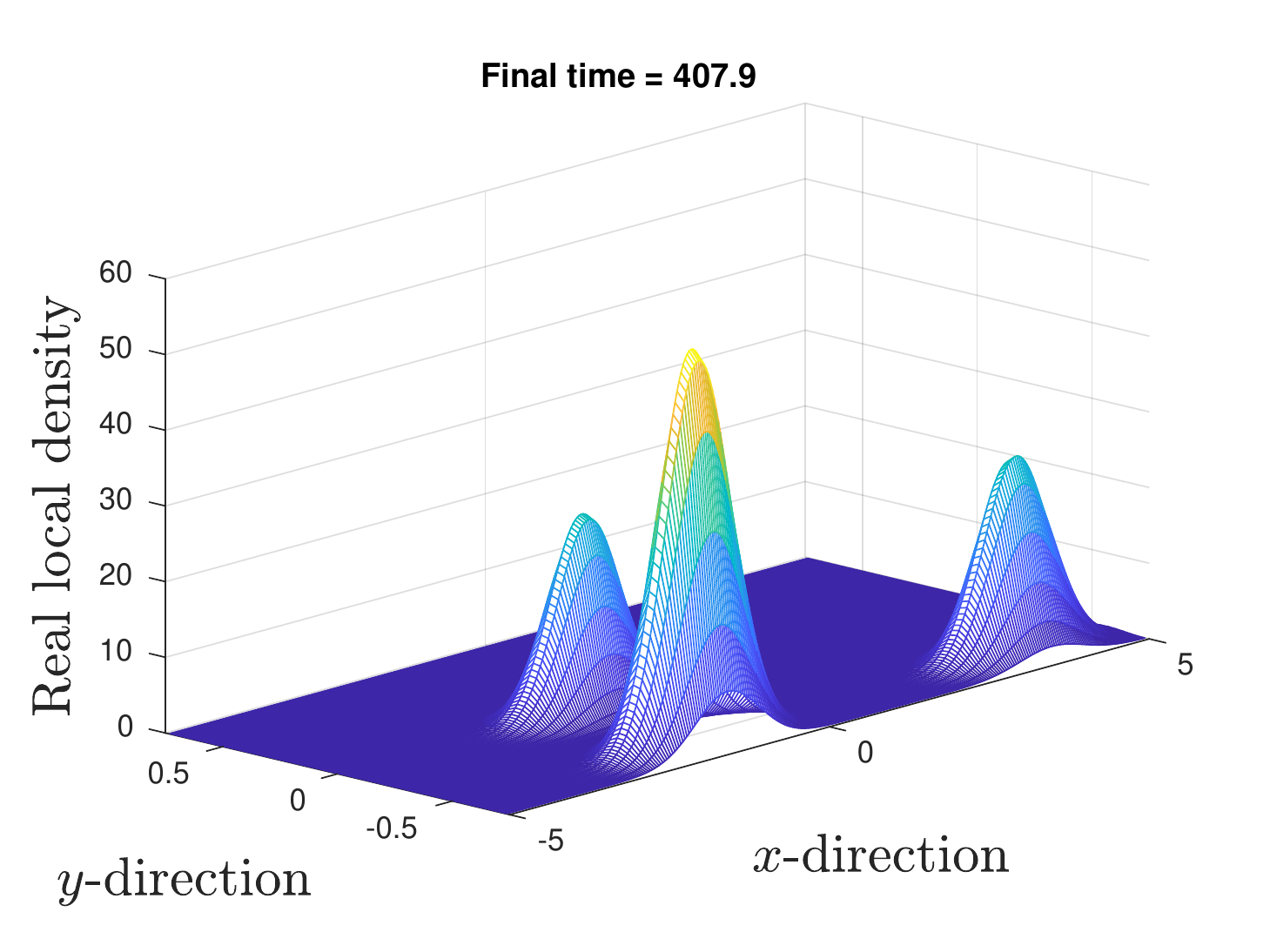}
	\includegraphics[width=0.49\textwidth]{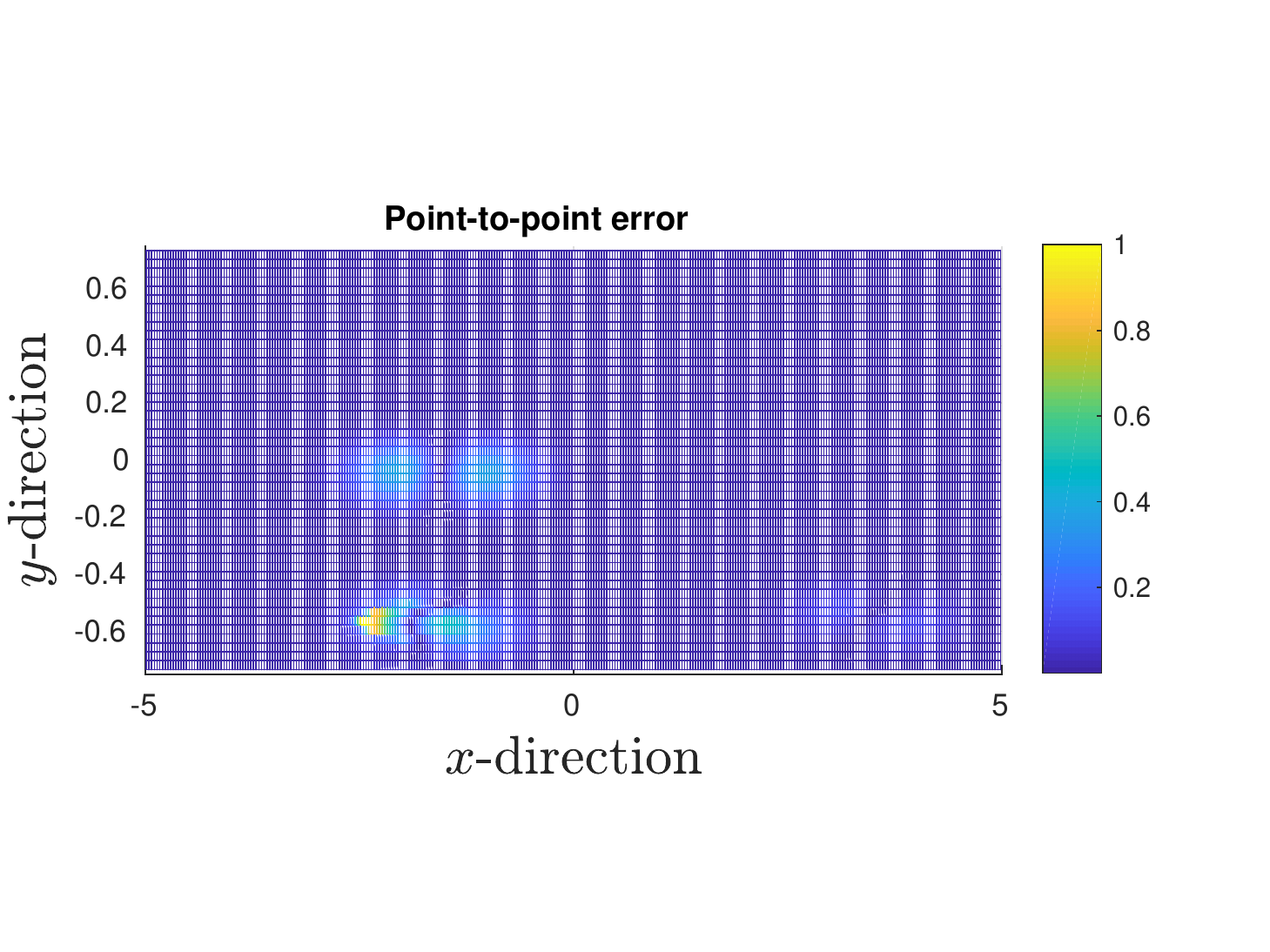}
	\caption{Top-left: initial density profile at time $T_{\text{init}} = 407.4$ seconds. Top-right: simulated density profile after $0.5$ seconds. Bottom-left: real density profile at final time. Bottom-right: difference between the simulated and the real density profiles.\label{fig:2DvsTrajectory1}}
\end{figure}

\begin{figure}[t!]
	\centering
	\includegraphics[width=0.49\textwidth]{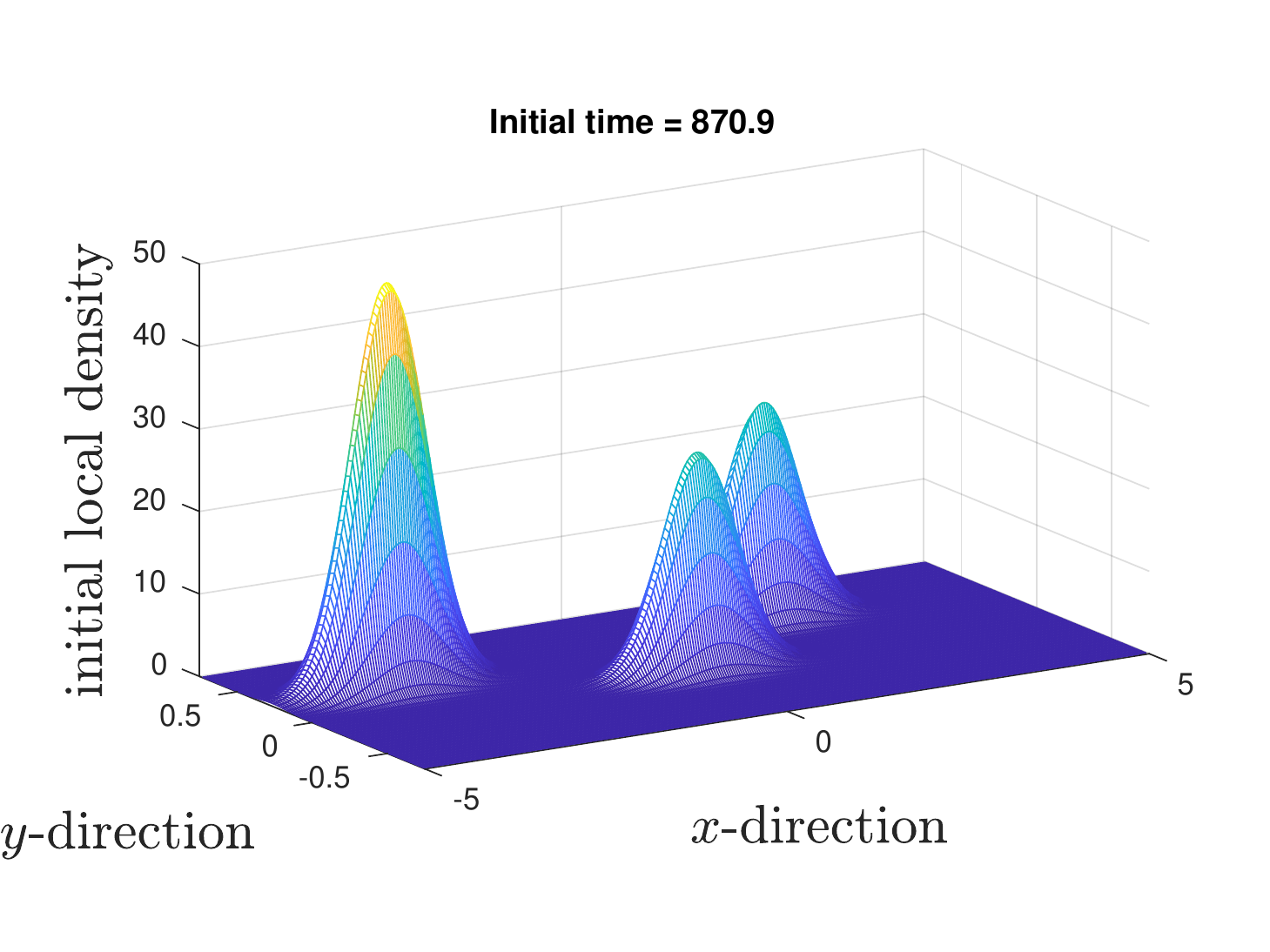}
	\includegraphics[width=0.49\textwidth]{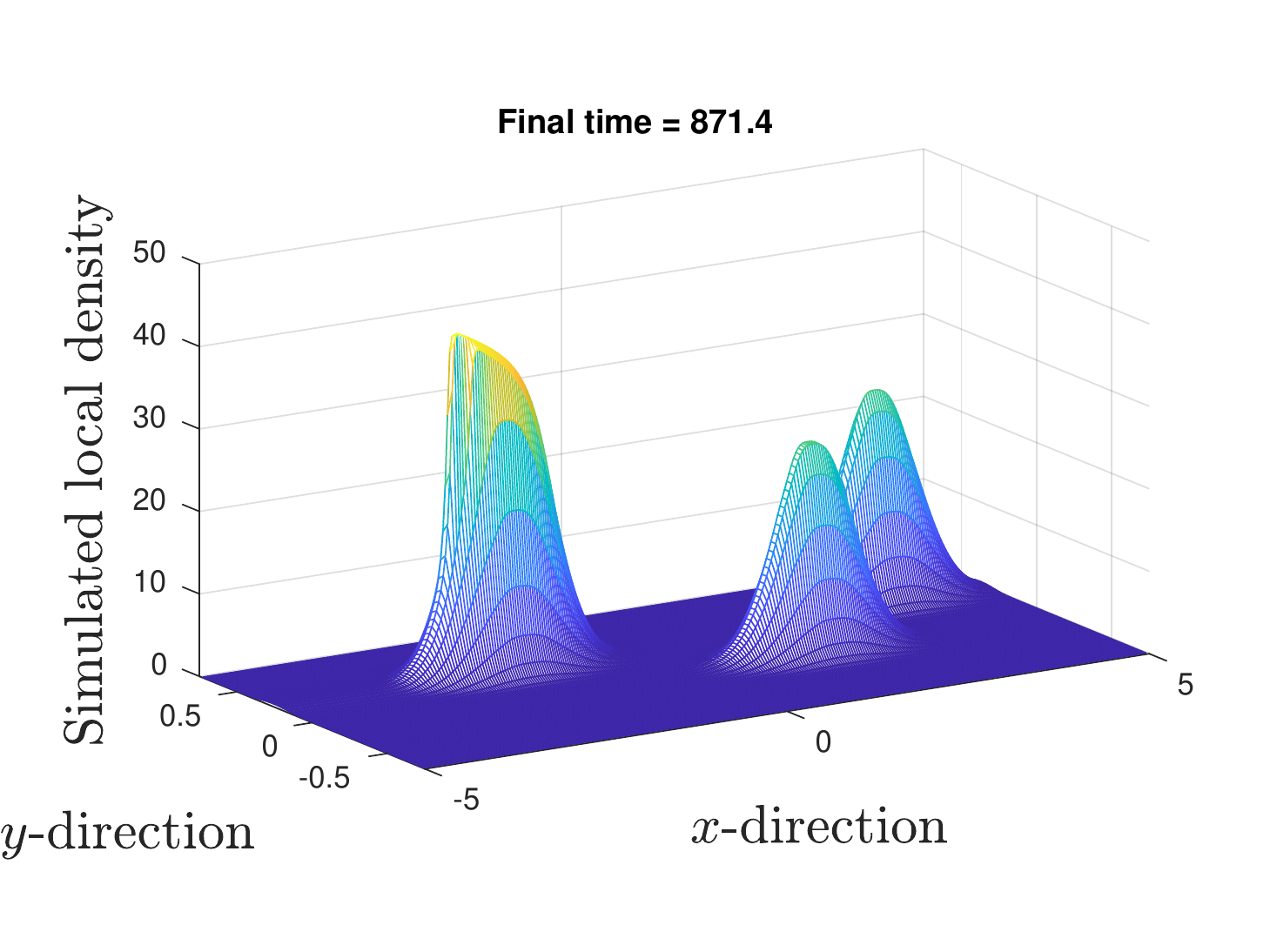}\\
	\includegraphics[width=0.49\textwidth]{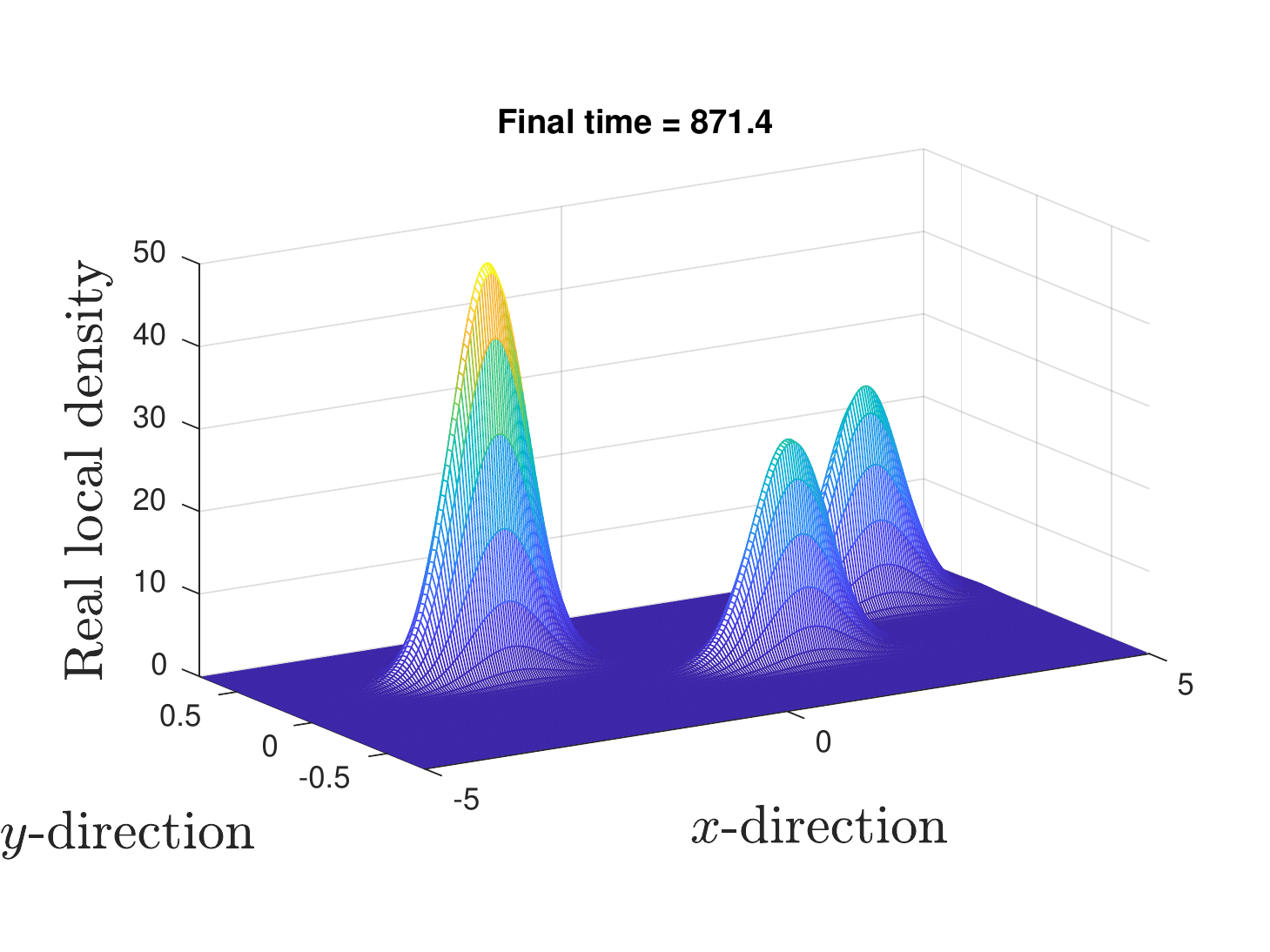}
	\includegraphics[width=0.49\textwidth]{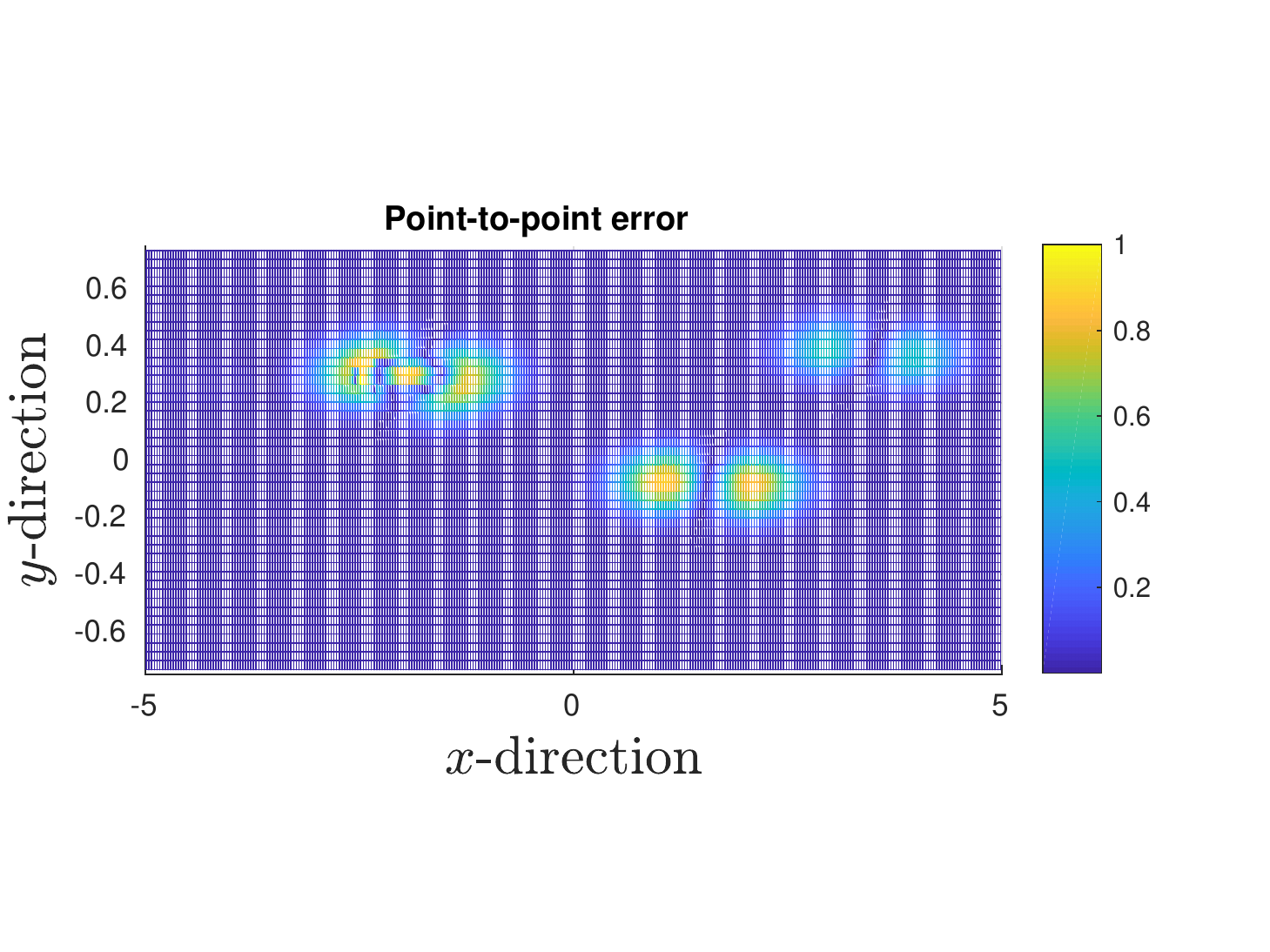}
	\caption{Top-left: initial density profile at time $T_{\text{init}} = 870.9$ seconds. Top-right: simulated density profile after $0.5$ seconds. Bottom-left: real density profile at final time. Bottom-right: difference between the simulated and the real density profiles.\label{fig:2DvsTrajectory2}}
\end{figure}

We study the predictive accuracy of the 2D model \eqref{eq:2DLWR} with respect to the trajectory data provided by the data. In the first test we simply study the accuracy of the model without possible spurious errors included by the treatment of boundary data. We choose an initial time $T_{\text{init}}$ and using the kernel density estimation approach we compute the density profile. Then we evolve it up to a final time $T_{\text{fin}}$, such that $T_{\text{fin}}-T_{\text{init}}= 0.5$ seconds, in order to guarantee that the simulation is not influenced by outgoing boundary conditions. Moreover, at the same time we wish to reduce errors due to the numerical scheme and therefore we choose a very small CFL condition. Finally, we compute the difference between the simulated profile and the real density profile at final time, normalizing with respect to the maximum value, c.f.  Figure \ref{fig:2DvsTrajectory1} and in Figure \ref{fig:2DvsTrajectory2}. 

Clearly, the boundary data are important for computing long time simulations. To this end, we extrapolate the incoming and the outgoing boundary data by artificially extending the trajectory data in computational cells outside the domain. In fact, recall that the trajectories of vehicles are approximated by means of a least squares linear approximation of their positions on the road section (see Section~\ref{sec:DataSet}), thus we are able to detect the cars in the ghost part of the computational domain at a fixed time. Then, using the KDE technique, we compute the two-dimensional density in the ghost cells which is used as boundary condition. More precisely, the extrapolation and the computation of the density in the ghost cells is based on the following procedure:
\begin{enumerate}
	\item as described in Section~\ref{sec:DataSet}, starting from the knowledge of the time dependent positions $(x_i(t),y_i(t))$ of each vehicle, we have considered the linear approximation of the data $\{t_k,x_i(t_k)\}_{k=m_i}^{M_i}$ and $\{t_k,y_i(t_k)\}_{k=m_i}^{M_i}$, where $t_{m_i}$ and $t_{M_i}$ would represent the minimum and the maximum time, respectively, such that $x_i(t) \in [0,L^x]$ for each $t\in[t_{m_i},t_{M_i}]$. Thus, for each vehicle $i$, we have the linear trajectories
	\begin{equation} \label{eq:linearapp}
	x(t) = m^x_i t + q^x_i, \quad y(t) = m^y_i t + q^y_i,
	\end{equation}
	minimizing
	$$
	\sum_{k=m_i}^{M_i} (x_i(t_k)-x(t_k))^2, \quad \sum_{k=m_i}^{M_i} (y_i(t_k)-y(t_k))^2.
	$$
	\item once the linear trajectories, and thus the coefficients $m^{x,y}_i$ and $q^{x,y}_i$ are known $\forall\,i$, we can use equations \eqref{eq:linearapp} to extrapolate the position along the road of a vehicle $i$ at time $t=\hat{t}$ such that $\hat{t}\notin[t_{m_i},t_{M_i}]$, computing
	$$
	x(\hat{t}) = m^x_i \hat{t} + q^x_i.
	$$
	\item using step 2, we count and identify the vehicles that at time $\hat{t}$ are in a position such that the KDE approach~\eqref{eq:DensityEstimation} applied to these data produces a nonzero density profile in the ghost cells. We use this density as boundary data. 
\end{enumerate}
Notice that, since we choose a very fine space discretization, the above extrapolation is supposed to be not too far beyond the known data.

\begin{figure}[t!]
	\centering
	\includegraphics[width=0.49\textwidth]{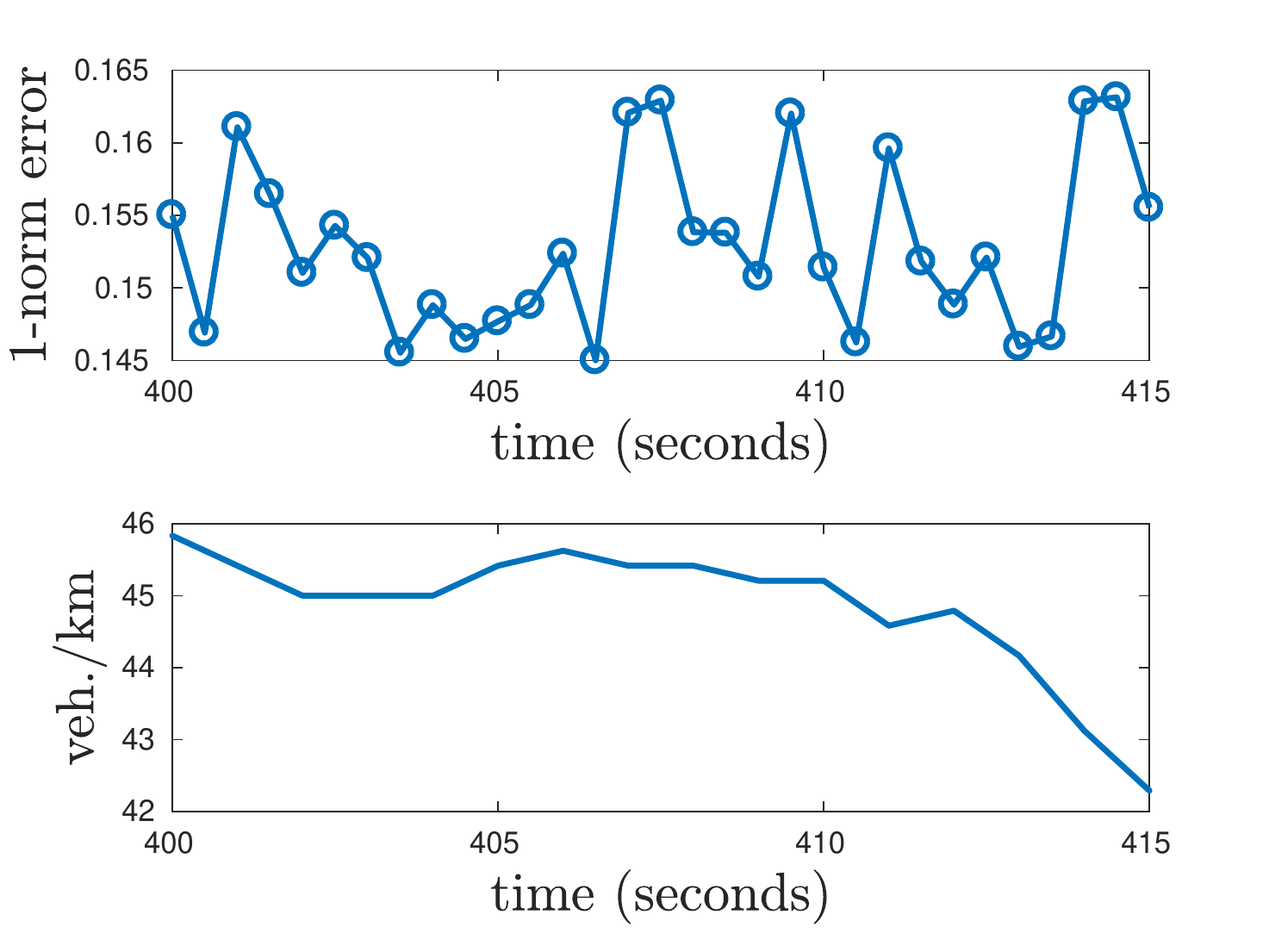}
	\includegraphics[width=0.49\textwidth]{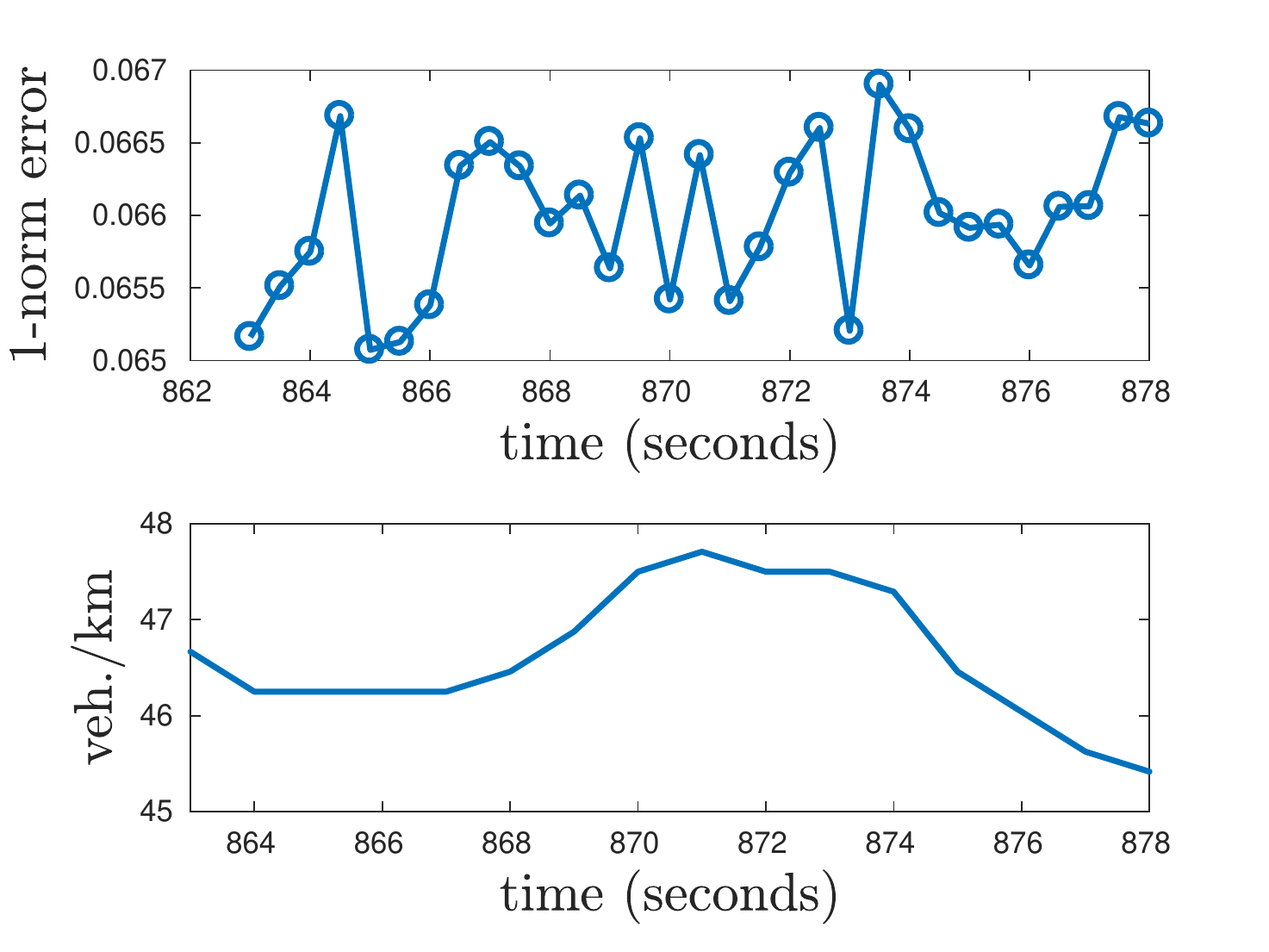}
	\caption{Left: $15$ seconds of simulation between $400-415$ seconds, showing the $1$-norm error each $0.5$ seconds (the top panel). Right: $15$ seconds of simulation between $863-878$ seconds, showing the $1$-norm error each $0.5$ seconds (the top panel). The mid and the bottom panels show the variation of density in the time intervals and on the whole recorded time period, respectively.\label{fig:LongTime}}
\end{figure}

In Figure \ref{fig:LongTime} we study the predictive accuracy of the 2D model \eqref{eq:2DLWR} for $15$ seconds taking into account the boundary data. We choose different time periods  for the simulations in Figure \ref{fig:2DvsTrajectory1} and in Figure \ref{fig:2DvsTrajectory2}. The error is computed every $0.5$ seconds on the whole domain using the $1$-norm error, see equation \eqref{eq:Error}. 

\subsubsection{Comparison between the 1D and the 2D model.}

We compare now the   2D model \eqref{eq:2DLWR} with respect its 1D version \eqref{eq:1DLWR} in order to estimate the benefit of a refined model compared with a commonly used averaged one-dimensional model. 

\begin{figure}[t!]
	\centering
	\includegraphics[width=0.49\textwidth]{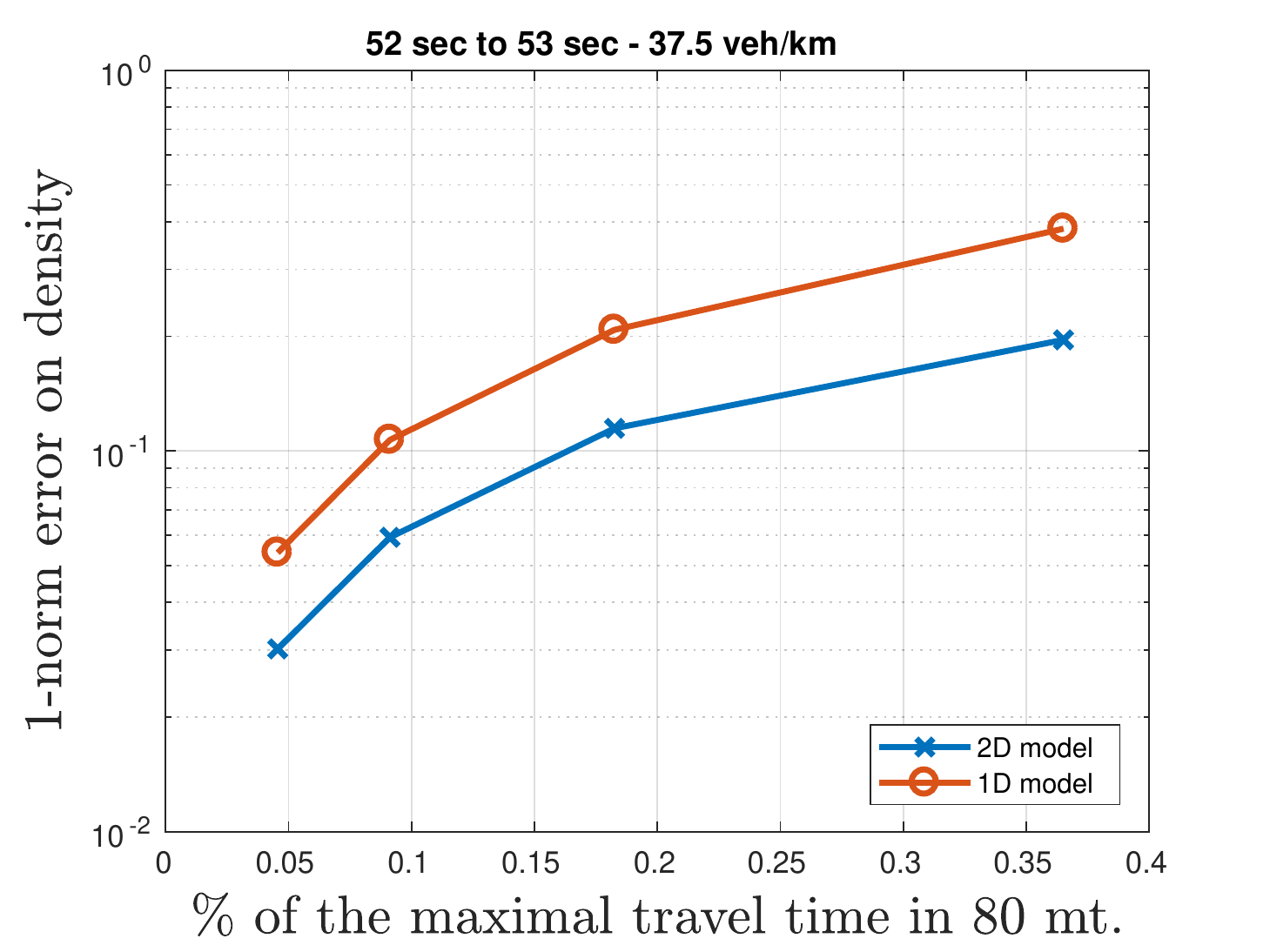}
	\includegraphics[width=0.49\textwidth]{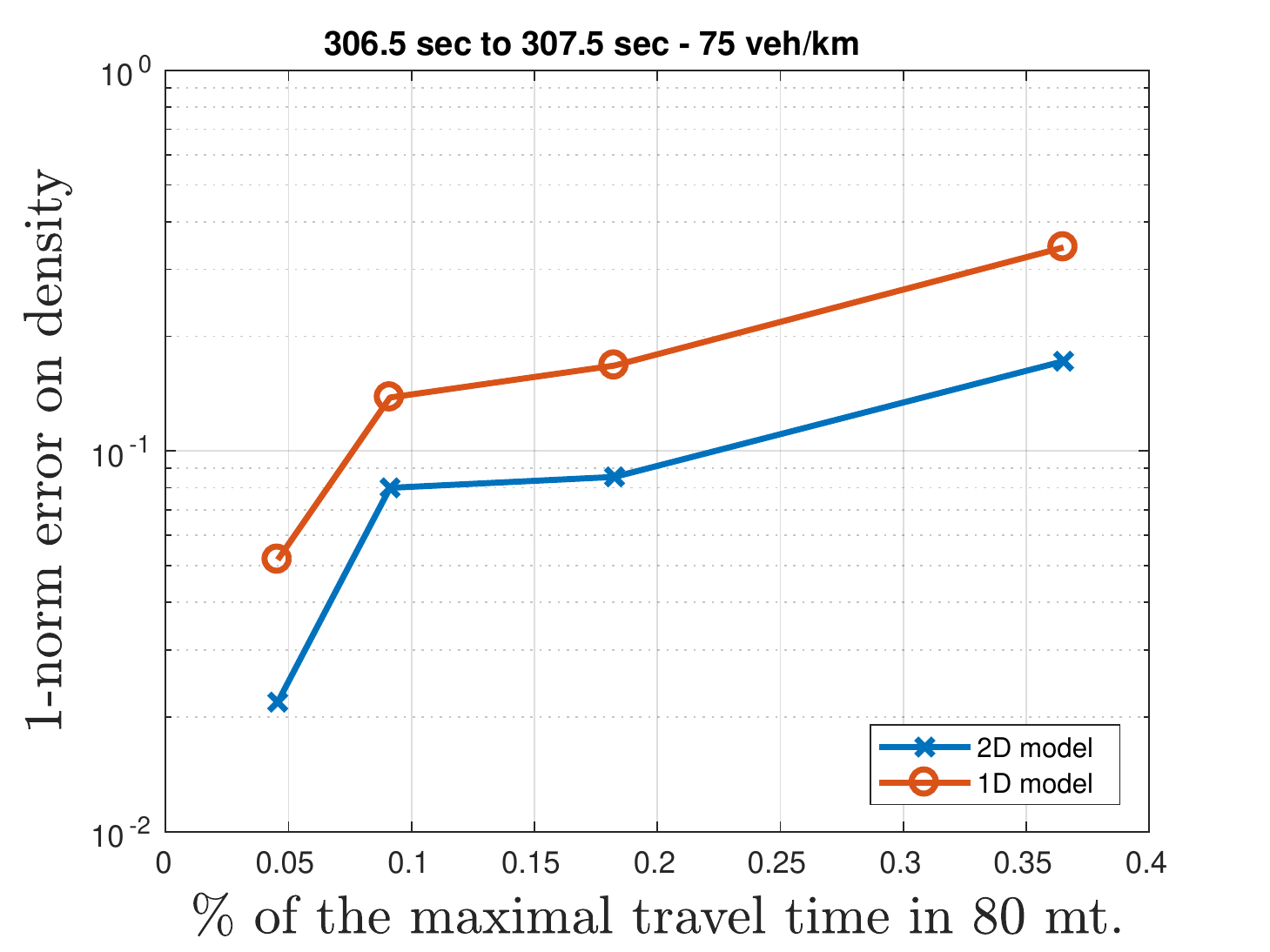}\\
	\includegraphics[width=0.49\textwidth]{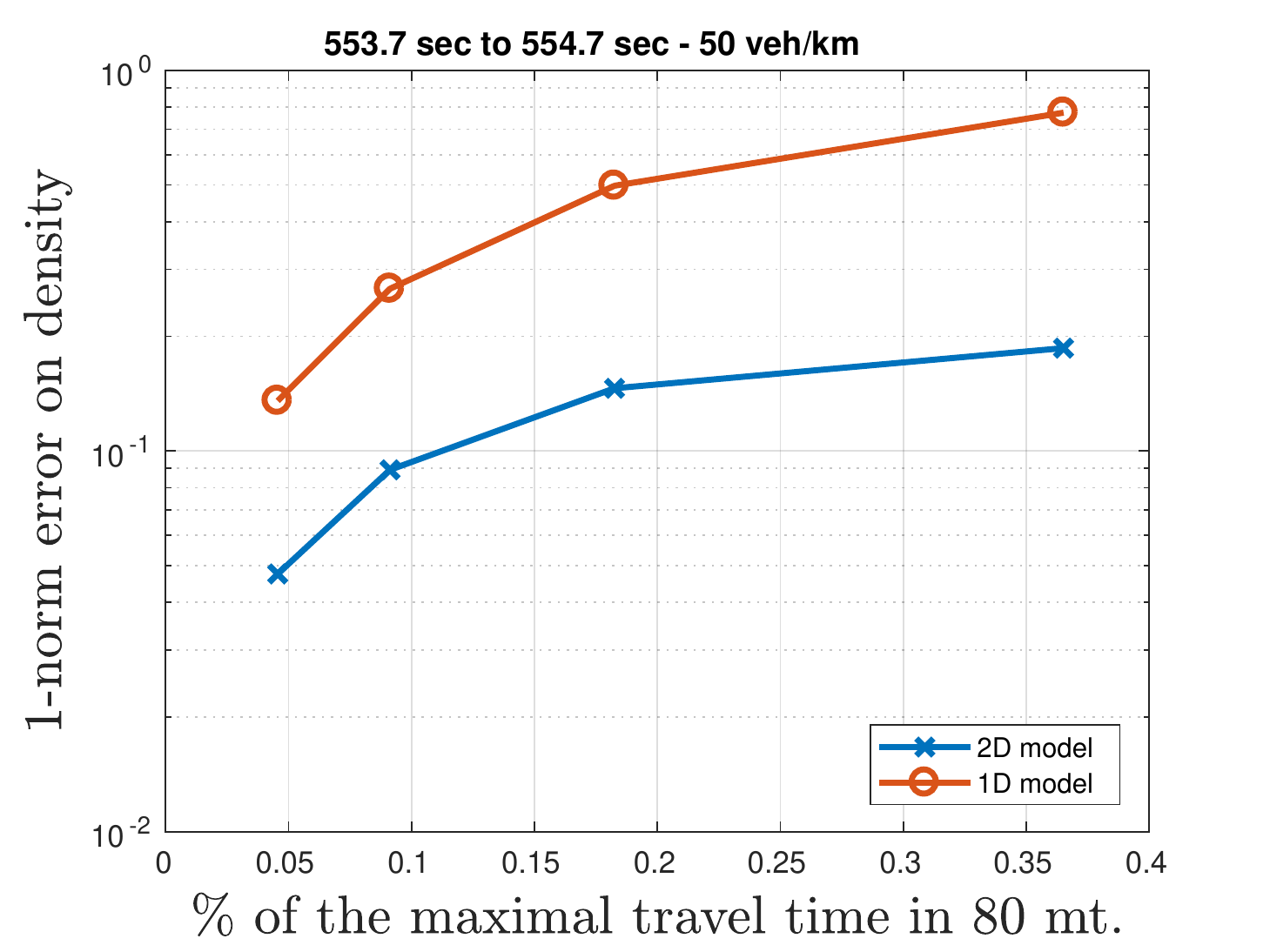}
	\includegraphics[width=0.49\textwidth]{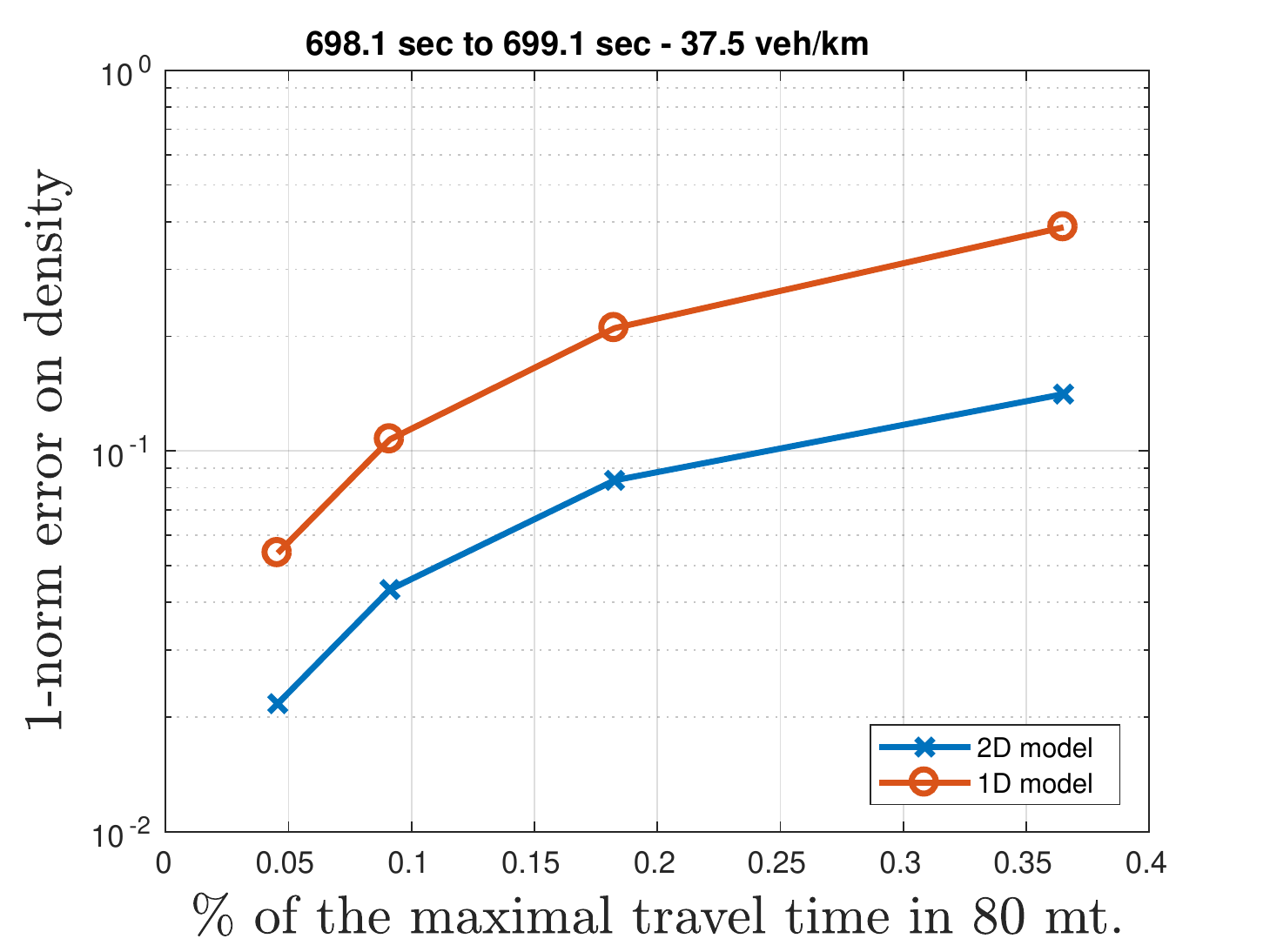}\\
	\includegraphics[width=0.49\textwidth]{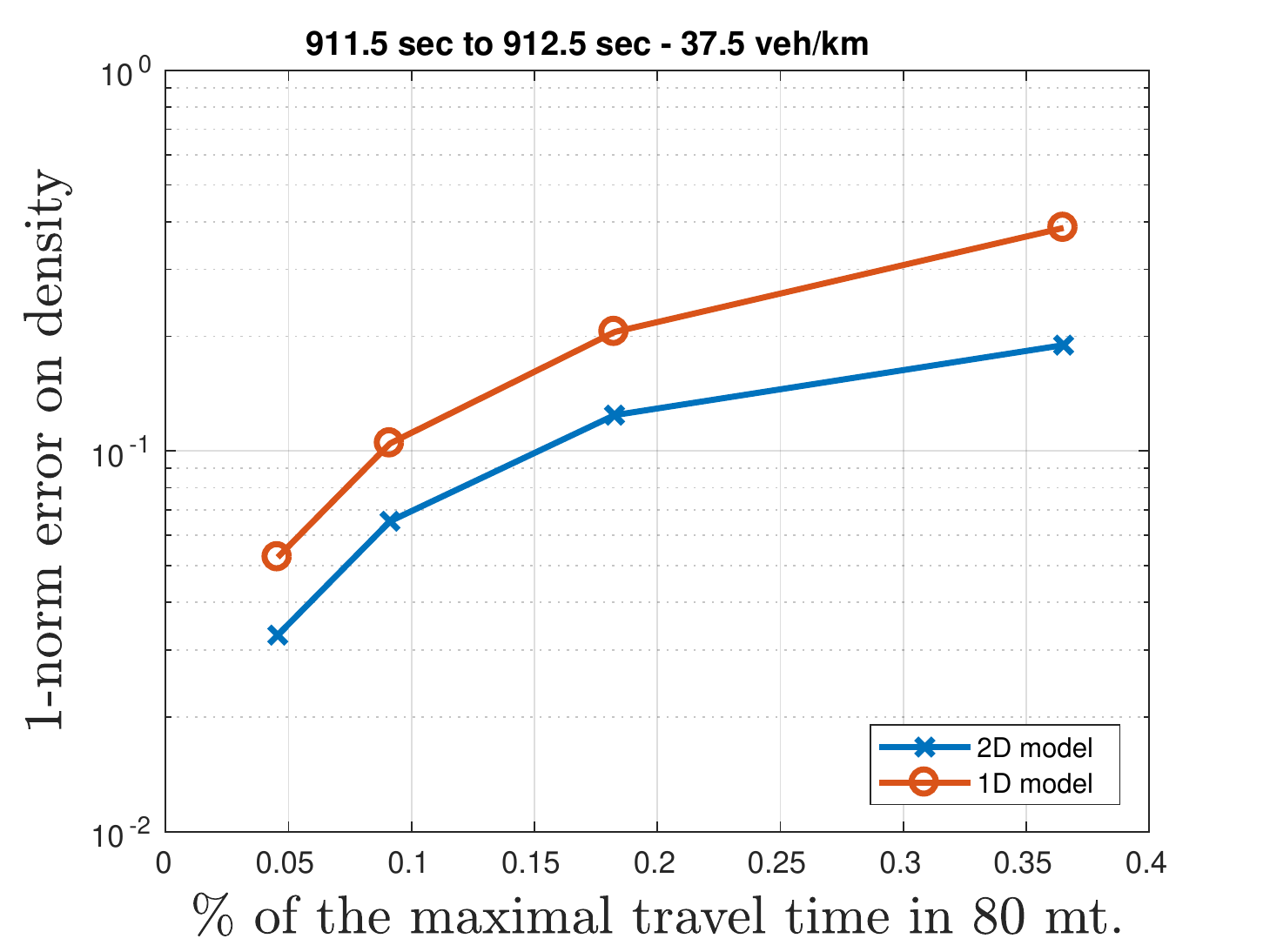}
	\includegraphics[width=0.49\textwidth]{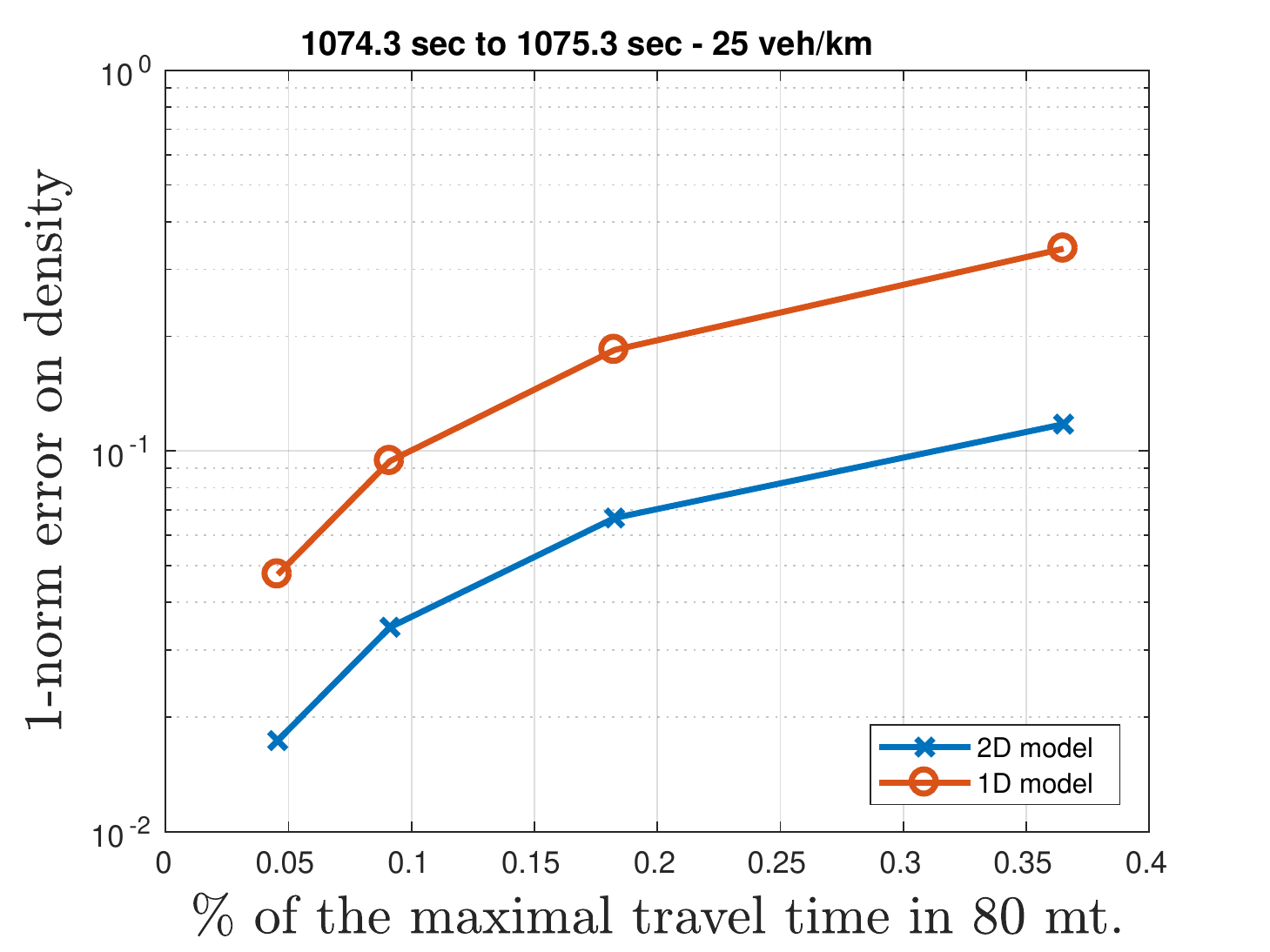}
	\caption{Error plots comparing the predictive accuracy of the 1D model \eqref{eq:1DLWR} (red data) and of the 2D model \eqref{eq:2DLWR} (blue data). Each panel refers to different initial density profiles computed by using the kernel density estimation approach. On the $x$-axis we show the percentage of the traveling time with respect to the total time to cover the road section at the maximum speed.\label{fig:1Dvs2D}}
\end{figure}

We select different initial conditions, characterized by different densities on the road. Then, we evolve the initial density profiles up to different final times $T_{\text{fin}} = 1/2^{i}$ seconds, with $i=0,1,2,3$.      	
In Figure \ref{fig:1Dvs2D} we compare the errors (the $1$-norm error as in \eqref{eq:Error}) on the density produced by the two macroscopic models \eqref{eq:1DLWR} (red data) and \eqref{eq:2DLWR} (blue data). The results show  that the 2D model produces smaller errors and therefore it can results in more realistic evolution of traffic conditions. This is mainly due to the fact that, for the 1D model \eqref{eq:1DLWR}, the kernel density estimation as well as the definition of the prediction error is done following the same approach of \cite{FanHertySeibold}. Hence, we project the $x$ positions on the same $y$ coordinate and this may result in an overestimation of the density. In the projected point of view, we could have two or more vehicles being near each other leading to, in the kernel density estimation approach, higher values of density traveling thus with a lower speed. But in the realistic data vehicles could be distant due to  different $y$ positions.

\begin{figure}[t!]
	\centering
	\includegraphics[width=0.49\textwidth]{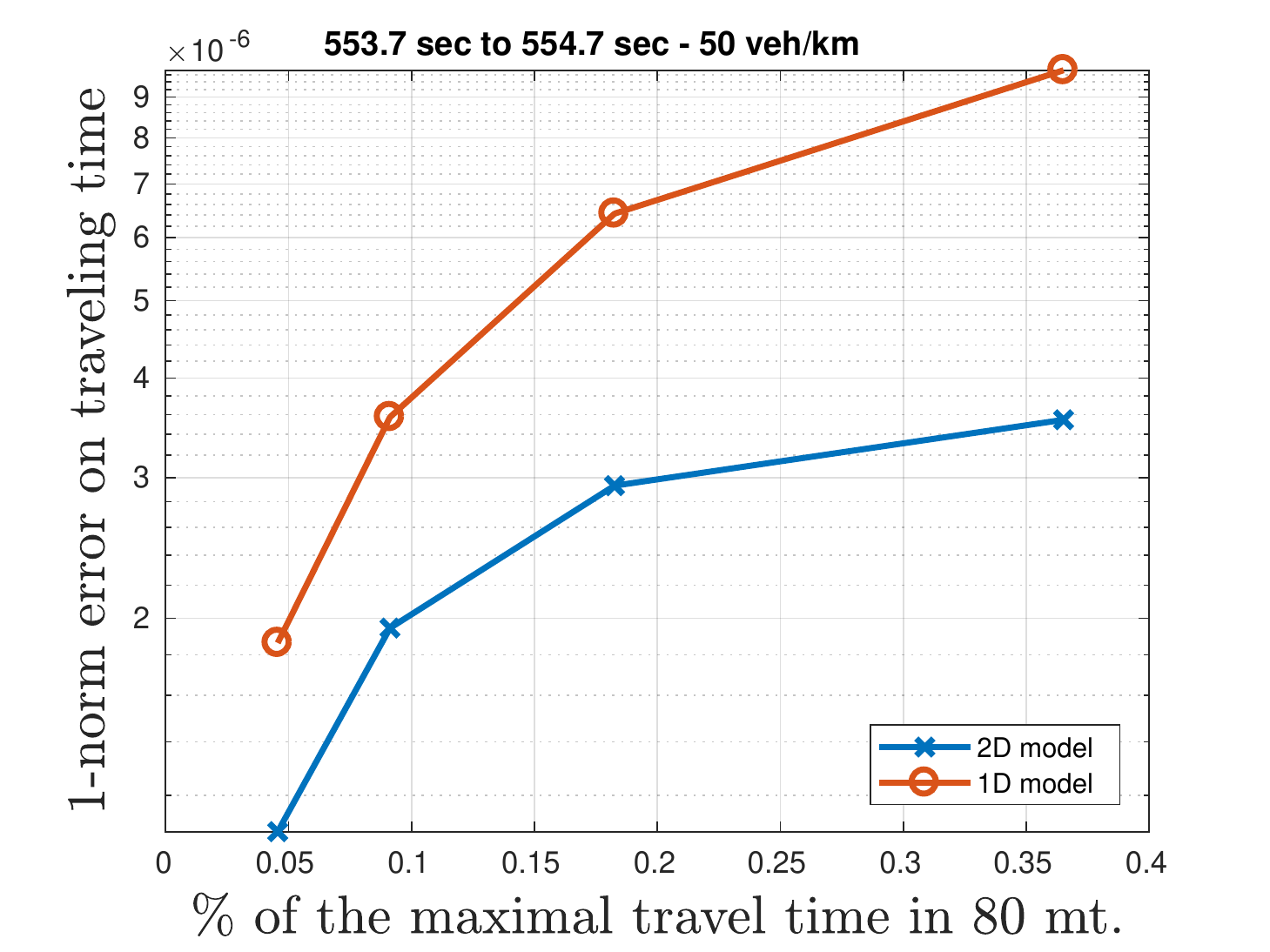}
	\caption{Error plots comparing the predictive prediction of traveling time of the 1D model \eqref{eq:1DLWR} (red data) and of the 2D model \eqref{eq:2DLWR} (blue data). On the $x$-axis we show the percentage of the traveling time with respect to the total time to cover the road section at the maximum speed.\label{fig:1Dvs2D-TT}}
\end{figure}

In order to better evaluate the performances between the 2D and the 1D model, in Figure~\ref{fig:1Dvs2D-TT} we also compare them in the prediction of traveling time, only for the case $T_{\text{init}}=553.7$, thus for the same initial time used in the middle left panel of Figure~\ref{fig:1Dvs2D}. The traveling time $TT$ for both models is computed up to the final times $T_{\text{fin}} = 1/2^{i}$ seconds, with $i=0,1,2,3$, as
$$
TT^{\text{1D}} = \frac{L^x}{u^x}, \quad TT^{\text{2D}} = \frac{L^x}{\sqrt{(u^x)^2+(u^y)^2}}
$$
where $u^x=q^x(\rho)/\rho$ and $u^y=q^y(\rho)/\rho$ are the speed functions computed using the flux functions $q^x$ and $q^y$ in equation~\eqref{eq:FitX} and equation~\eqref{eq:FitY}, respectively, with the optimal parameters determined in Section~\ref{sec:DataFitting}. The exact traveling time is instead computed by applying the kernel density estimation approach to find a continuous estimation for the fluxes at time $t$ as
\begin{align*}
\text{Flux}^x(t,x,y) = \sum_{i=1}^{N(t)} v_i^x K\big(x-x_i(t),y-y_i(t)\big)\\
\text{Flux}^y(t,x,y) = \sum_{i=1}^{N(t)} v_i^y K\big(x-x_i(t),y-y_i(t)\big),
\end{align*}
where $K$ is defined in~\eqref{eq:K} and $v_i^x$, $v_i^y$ are the microscopic velocities in $x$- and $y$-direction of vehicle $i$ being on the road section at time $t$. Finally, a continuous estimation for the speeds is computed as
$$
\text{Speed}^x(t,x,y) = \frac{\text{Flux}^x(t,x,y)}{\rho(t,x,y)}, \quad \text{Speed}^y(t,x,y) = \frac{\text{Flux}^y(t,x,y)}{\rho(t,x,y)}
$$
where $\rho$ is given by~\eqref{eq:DensityEstimation} and the exact traveling time, for the 2D case, is
$$
TT = \frac{L^x}{\sqrt{(\text{Speed}^x)^2+(\text{Speed}^y)^2}}.
$$
The traveling time for the 1D model is instead computed by applying the kernel density estimation approach for the flux following the same approach of~\cite{FanHertySeibold} and thus, as done above, by projecting the $x$ positions on the same $y$ coordinate. In this way, we get the estimation for the flux in $x$-direction and then we can compute the estimation for the velocity which is used to find the exact traveling time.

\subsubsection{Dependence on the fitting parameters.} We study the dependence of the presented results on the choice of the fitting parameters, those being crucial in the derivation. In particular, we are interested in the magnitude of the error changes due to the variation of the parameters defining the closure in $y$-direction, see equation \eqref{eq:FitY}. 

\begin{figure}[t!]
	\centering
	\includegraphics[width=0.49\textwidth]{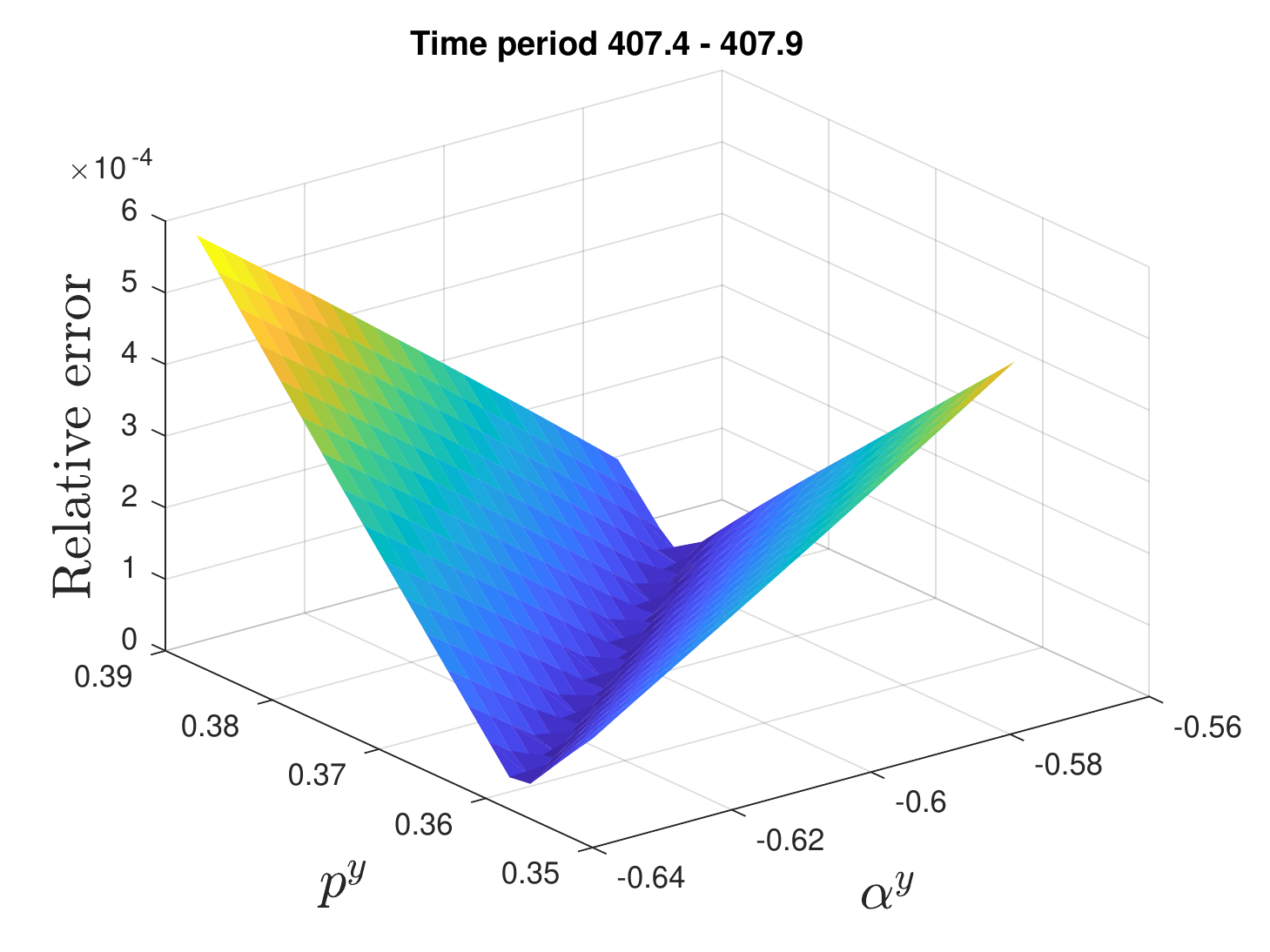}
	\includegraphics[width=0.49\textwidth]{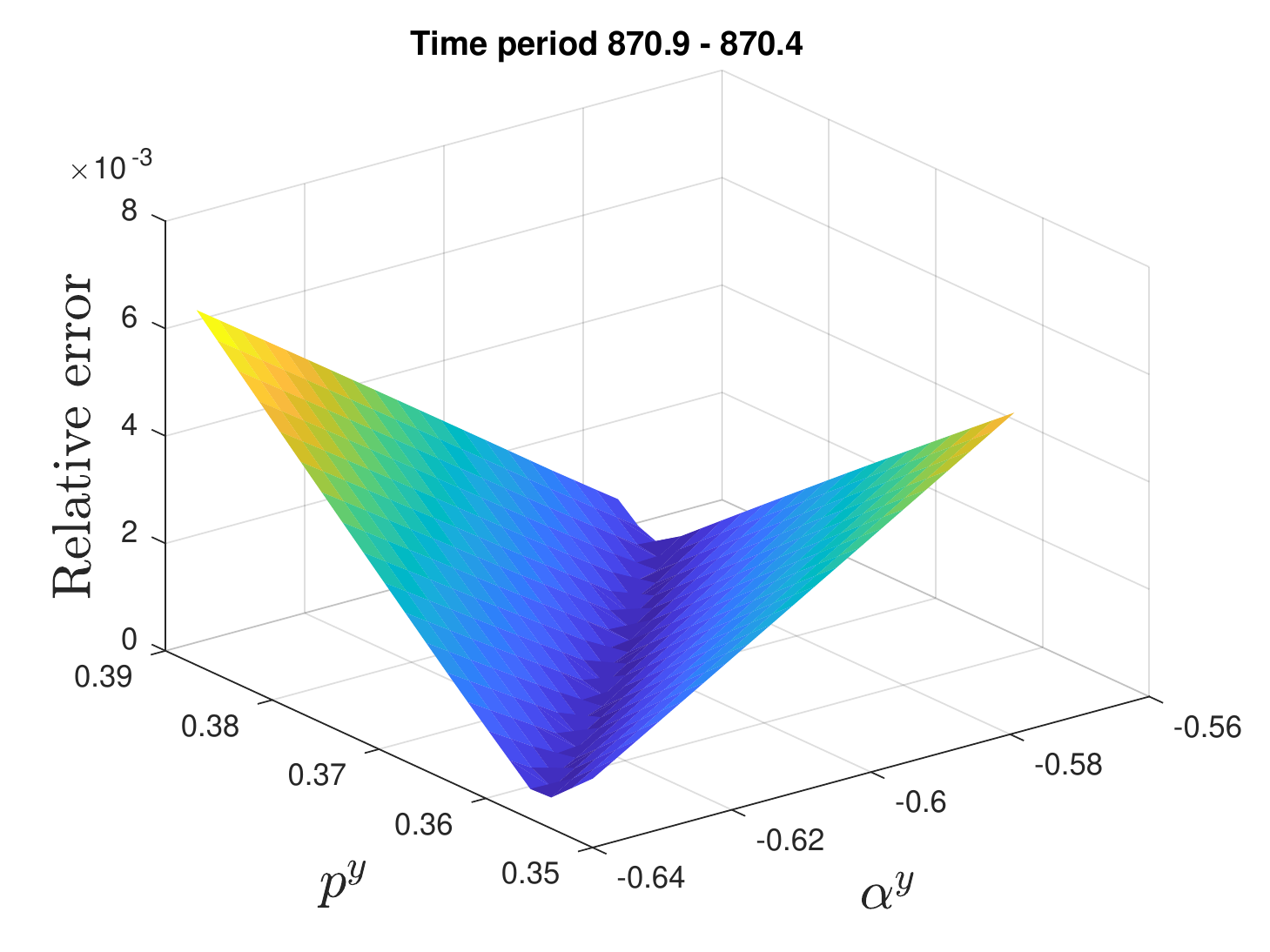}\\
	\caption{Relative error between $\boldsymbol{E}_{\boldsymbol{\alpha}^{\boldsymbol{y}},\boldsymbol{p}^{\boldsymbol{y}}}$ and $E_{\alpha^y_{\text{opt}},p^y_{\text{opt}}}(T_{\text{fin}})$ for each pair of possible parameters in the vectors $\boldsymbol{\alpha}^{\boldsymbol{y}}$ and $\boldsymbol{p}^{\boldsymbol{y}}$.\label{fig:RelativeError}}
\end{figure}

We consider the same initial conditions as  studied in Figure \ref{fig:2DvsTrajectory1} and in Figure \ref{fig:2DvsTrajectory2}. In both cases we consider $20$ values of the fitting parameters $\alpha^y$ and $p^y$ sampled from the intervals $\boldsymbol{\alpha}^{\boldsymbol{y}}=\left[\alpha^y_{\text{opt}}(1+5\%),\alpha^y_{\text{opt}}(1-5\%)\right]$ and $\boldsymbol{p}^{\boldsymbol{y}}=\left[p^y_{\text{opt}}(1-5\%),p^y_{\text{opt}}(1+5\%)\right]$, respectively. Then, we evolve the density profile with the 2D model \eqref{eq:2DLWR} for $0.5$ seconds and for each of the $400$ pairs $(\alpha^y,p^y)$. Finally, we compute the errors $\boldsymbol{E}_{\boldsymbol{\alpha}^{\boldsymbol{y}},\boldsymbol{p}^{\boldsymbol{y}}}$ at final time as in \eqref{eq:Error}. %In Table \ref{tab:Errors} we show the errors obtained with some of the previous pairs $(\alpha^y,p^y)$.
We recall that the error for the optimal pair $(\alpha^y_{\text{opt}},p^y_{\text{opt}})$ is $E_{\alpha^y_{\text{opt}},p^y_{\text{opt}}}(T_{\text{fin}})=0.1540$ for the time period $407.4-407.9$ and $E_{\alpha^y_{\text{opt}},p^y_{\text{opt}}}(T_{\text{fin}})=0.0660$ for the time period $870.9-871.4$. Notice that the different parameters do not modify the errors strongly and therefore the presented procedure is robust against those variations. In order to quantify this consideration, in Figure \ref{fig:RelativeError} we show the relative difference between $\boldsymbol{E}_{\boldsymbol{\alpha}^{\boldsymbol{y}},\boldsymbol{p}^{\boldsymbol{y}}}$ and $E_{\alpha^y_{\text{opt}},p^y_{\text{opt}}}(T_{\text{fin}})$. We observe that the maximum differences are of  order $10^{-4}$ and $10^{-3}$ and therefore of the order of the numerical scheme.

\section{Conclusions and Outlook} \label{sec:Conclusions}

In this paper we proposed a two-dimensional scalar macroscopic model to describe traffic flow on multi-lane roads. Therefore, the equation generalizes the one-dimensional LWR model. We prescribed the closure laws describing the two flux functions by using a data-fitting technique with respect to experimental measurements on a German highway. 

\begin{figure}[t!]
	\centering
	\includegraphics[width=0.49\textwidth]{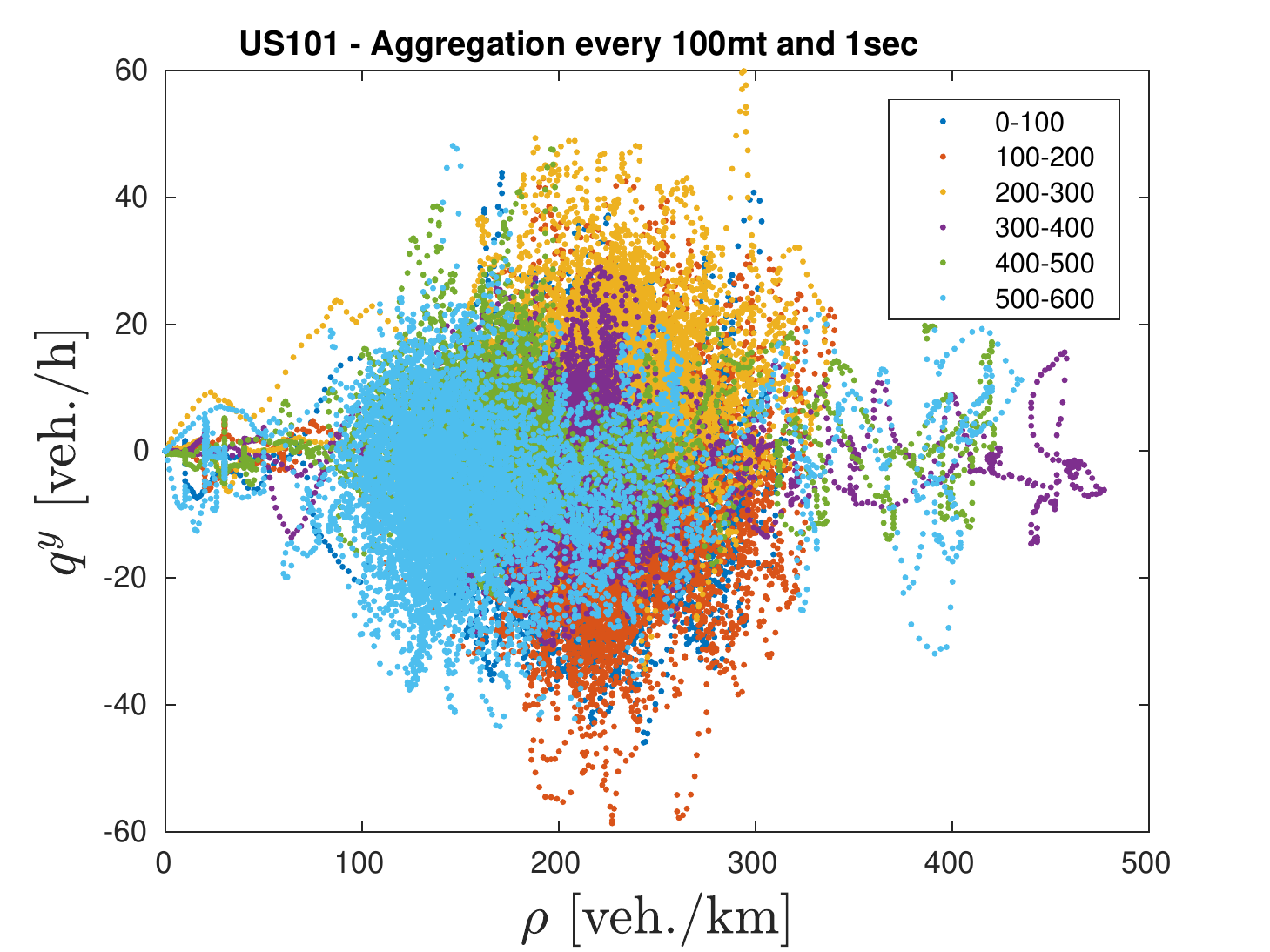}
	\includegraphics[width=0.49\textwidth]{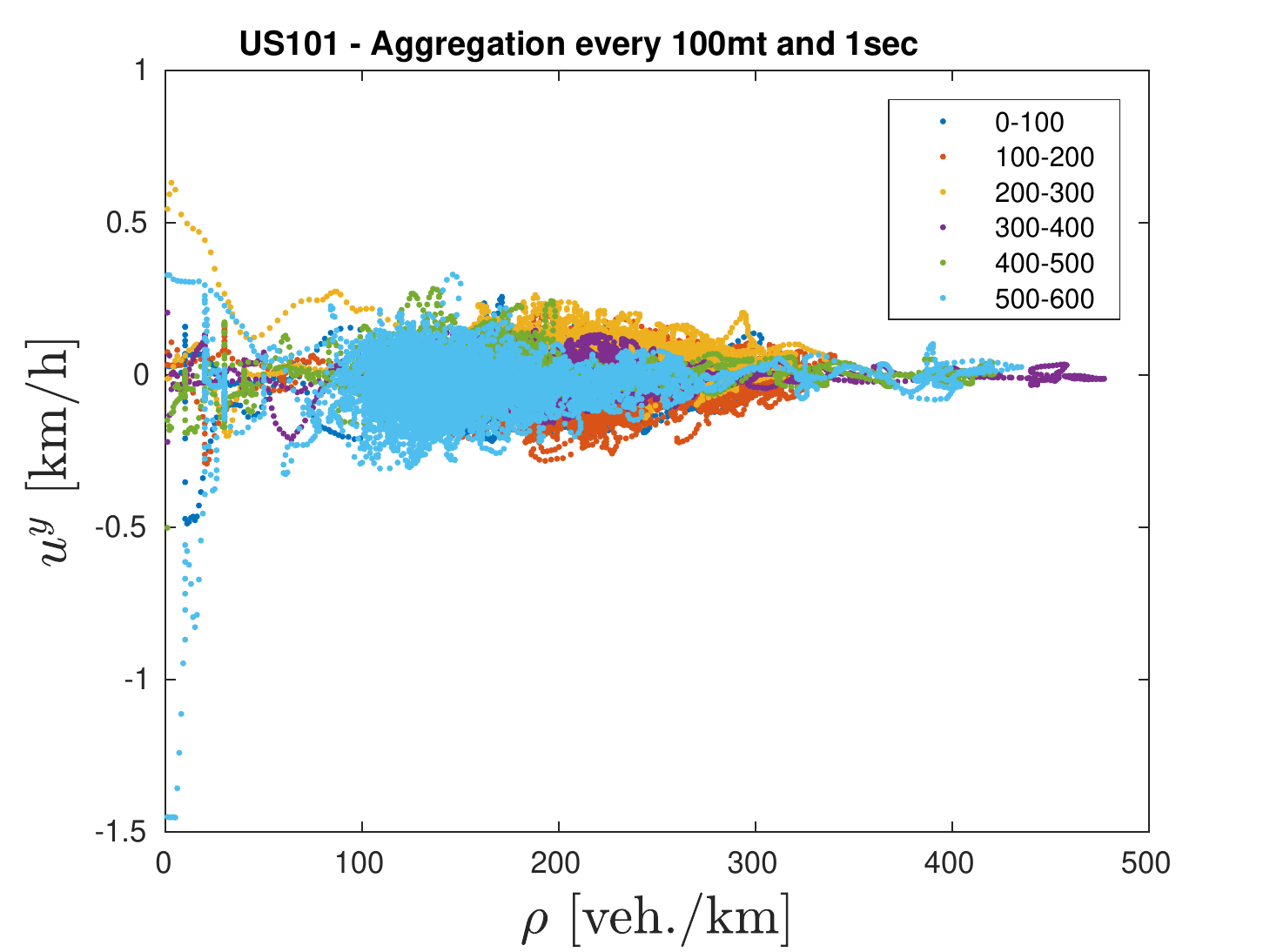}
	\caption{Experimental diagrams in $y$-direction resulting from the US101 highway and using $15$~minutes of recorded data (07:50 - 08:05 a.m.). The macroscopic quantities are obtained by aggregating each $100$ meter sections and every $1$ second. Left panel: flux-density diagram. Right: speed-density diagram.\label{fig:USdata}}
\end{figure}

Since laser sensors provide the two-dimensional trajectory of vehicles, we recovered also the fundamental diagram for traffic behavior across lanes. To our knowledge, this the first time that the resulting behavior of the flow across lanes is taken into account. This is possible thanks to the particular traffic rules on European highways which lead to a non-naive dynamics in the orthogonal direction to the movement of vehicles. In fact, if we consider experimental data on US highway, where there is no obligation to overtake on left lanes, the resulting behavior across lanes is naive. For instance, see Figure \ref{fig:USdata} in which we show the fundamental and speed-density diagrams in $y$-direction computed from NGSIM data on US101~\cite{NGSIM}. The mean dynamics of the flow seems to suggest $q^y = 0$ and therefore the 2D model~\eqref{eq:2DLWR} reduces to the 1D LWR-type model \eqref{eq:1DLWR}.

On the German data-set, numerical examples show the validity of the macroscopic modeling when comparing with experiments. In particular, the numerical comparison with trajectory data shows that the two-dimensional scalar model already outperforms a corresponding lane-averaged one-dimensional model. From an application point of view, in future works we plan to investigate now the effect of regulations on lane reduction using the two-dimensional setting as well as higher-order models. In fact, it is expected that as in \cite{FanHertySeibold} the additional degree of freedom in the modeling allows for a better adjustment of the models to data. Further, in~\cite{HertyMoutariVisconti} we study a second-order two-dimensional macroscopic model as limit of a two-dimensional microscopic follow-the-leader model which is validated using the German data-set. Instead, in~\cite{HertyTosinViscontiZanella} we study a two-dimensional hybrid kinetic model which allows to take into account the two-dimensional phenomena of traffic flow in the kinetic theory showing, as application, that it is able to reproduce the US data-sets.

%For acknowledgements section, please don't number the section, please begin it with \section*{Acknowledgements}
\section*{Acknowledgments} This work has been supported by HE5386/13-15 and DAAD MIUR project.
We also thank the ISAC institute at RWTH Aachen, Prof. M. Oeser and MSc. F. Hennecke
for kindly providing the trajectory data.

\bibliographystyle{plain}
\bibliography{completeBibTex,references}

\end{document}